\documentclass{article}
\usepackage[utf8]{inputenc}
\usepackage{cite}
\usepackage{graphicx}
\usepackage{amsmath}
\usepackage{amssymb}
\usepackage{algorithm}
\usepackage{algorithmic}
\usepackage{appendix}
\usepackage{color}
\usepackage[margin = 3cm]{geometry}
\usepackage{hyperref}
\usepackage{DOTacronyms}
\usepackage{lineno}

\newcommand{\Changes}[1]{\textcolor{black}{#1}}

\newcommand{\address}[1]{\vspace{1cm}\\#1}
\newcommand{\authormark}[1]{$^{#1}$}
\DeclareMathOperator*{\E}{\mathbb{E}}
\DeclareMathOperator*{\argminA}{arg\,min} 
\newcommand{\norm}[1]{\left\lVert#1\right\rVert}

\newcommand{\eref}[1]{Eq. (\ref{#1})} 

\newcommand{\pos}{r}
\newcommand{\ang}{s}
\newcommand{\angvec}{\hat{{\bf \ang}}}
\newcommand{\domain}{\Omega}
\newcommand{\bdomain}{\partial \domain}
\newcommand{\rmd}{\mathrm{d} }
\newcommand{\MC}{
\ensuremath {\mathcal{L}}^{-1}_{\rm MC}
}
\newcommand{\AMC}{ 
\ensuremath {\mathcal{L}}^{-1\ast}_{\rm MC} 
}
\newcommand{\Obj}{ F}
\newcommand{\qObj}{\Obj^{\rm QPAT} }
\newcommand{\uObj}{\Obj^{\rm UMOT} }
\newcommand{\mua}{\mu_{\rm a}}
\newcommand{\mus}{\mu_{\rm s}}
\newcommand{\ydata}{y}
\newcommand{\yobs}{\ydata^{\rm obs}}
\newcommand{\udata}{b}
\newcommand{\uobs}{\udata^{\rm obs}}
\newcommand{\qdata}{h}
\newcommand{\qobs}{\qdata^{\rm obs}}
\newcommand{\Frechet}{Fr{\'{e}}chet}
\newcommand{\dontshow}[1]{}

\title{Efficient inversion strategies for estimating optical properties with Monte Carlo radiative transport models}

\author{Callum M. Macdonald\authormark{1}, Simon Arridge\authormark{2}, Samuel Powell\authormark{3}\\
\address{
\authormark{1} \footnotesize{Department of Medical Physics and Biomedical Engineering, University College London, London, WC1E 6BT, UK.}\\
\authormark{2} \footnotesize{Department of Computer Science, University College London, London, WC1E 6BT, UK.}\\
\authormark{3} \footnotesize{Faculty of Engineering, University of Nottingham, Nottingham, NG7 2RD, UK.}\qquad
}
}

\date{\today}

\let\vec\mathbf

\begin{document}

\maketitle

\begin{abstract}

\noindent{\bf{Significance:}} 
Indirect imaging problems in biomedical optics generally require repeated evaluation of forward models of radiative transport, for which Monte Carlo is accurate yet  computationally costly.  We develop a novel approach to reduce this bottleneck which has significant implications for quantitative tomographic imaging in a variety of medical and industrial applications.\\

\noindent{\bf{Aim:}} 
Our aim is to enable computationally efficient image reconstruction in (hybrid) diffuse optical modalities using stochastic forward models.\\

\noindent{\bf{Approach:}} 
Using Monte Carlo we compute a fully stochastic gradient of an objective function for a given imaging problem. Leveraging techniques from the machine learning community we then adaptively control the accuracy of this gradient throughout the iterative inversion scheme, in order to substantially reduce computational resources at each step.\\

\noindent{\bf{Results:}} 
For example problems of \acl{QPAT} and \acl{UMOT}, we demonstrate that solutions are attainable using a total computational expense that is comparable to (or less than) that which is required for a \textit{single} high accuracy forward run of the same Monte Carlo model.\\

\noindent{\bf{Conclusions:}} 
This approach demonstrates significant computational savings when approaching the full non-linear inverse problem of optical property estimation using stochastic methods.  \\


\end{abstract}
\bigskip

\section{Introduction}
\label{sec:intro}

Inverse problems arise in many areas within biomedical optics, both for global characterisation of optical properties of media and for image reconstruction,  amongst other applications\cite{arridge2009a}.
Inverse problems are often considered as optimisation problems, solved by deriving the gradient of an objective function and iteratively descending through the solution space. This process requires repeated solutions of forward and corresponding adjoint problems which are often computationally demanding in their own right. If the forward problem is given by the solution to a \ac{PDE} then one appealing approach is to solve the forward and inverse problems \emph{simultaneously} so that at intermediate stages in the algorithm (i.e. before it has finally converged) the forward problem is only approximately solved; this approach (which has its basis in Optimal Control) is known as \emph{PDE-constrained Optimisation}~\cite{biegler2003,bredies2013,reyes2015}. In this work we seek an equivalent framework for the case where approximate noisy solutions to the forward model can (or must) be sought by stochastic methods. 

The application of stochastic methods for the solution of \ac{PDE}s is particularly pertinent in problems involving diffuse optics, since the ``gold standard'' method of solving the \ac{RTE} - which is the most generally applicable description of the the underlying physics - is to use stochastic (Monte Carlo) techniques~\cite{zhu2013review}; their use in such applications parallels their extensive employment in other fields such as Neutron Physics~\cite{lux1991}. Whilst approximations to the \ac{RTE} 
(such as diffusion)  which permit deterministic solutions are available, these are often not valid in many cases such as in small domains, close to sources and boundaries, and in regions with weak scattering or strong absorption. 
\Changes{Analytical solutions to the \ac{RTE} itself are known for some geometries, such as infinite space\cite{LiemertKienle2011}, and layered media\cite{LiemertKienle2017}, but such expressions are not readily available for general domains}.  
The practicality of Monte Carlo techniques has been significantly boosted by recent advances in computational hardware developments, particularly in the application of parallelization~\cite{lo2009hardware,ren2010gpu}. Other approaches to improve their \Changes{computational} performance have been explored, such as the introduction of perturbation techniques~\cite{hayakawa2001perturbation}, or variance reduction techniques~\cite{Graff93,lima2011improved}. Consequently, even when \Changes{the} aforementioned approximations to the RTE are reasonable, Monte Carlo solutions may offer an attractive alternative to the use of deterministic techniques such as the finite element method, when the complexity of the geometry or probe requires a high density discretisation of the spatial domain.

With both deterministic and stochastic solvers, the computational cost of the forward model typically remains the limiting factor in image reconstruction procedures. However stochastic methods have a particular quality distinct from deterministic methods: one may arbitrarily trade computational expense against noise in the estimated solution, without bias. In the case of diffuse optics, this trade off is mediated through the number of virtual photons simulated by the Monte Carlo model for a given problem. This fact naturally leads one to consider how much noise can be tolerated during the solution of the inverse problem, and if a strategy can be found by which to approach this solution with the least work. 

Parallels can be drawn between this problem and large scale machine learning, where the requirement is to find the global minimum of a loss function expressed through fitting a model to a very large set of training data. The recent growth in this field has led to significant developments in optimisation methods using \emph{stochastic subsets}, and in particular the use of \emph{approximate gradients} at intermediate steps, a technique known as \acf{SGD}. At the heart of this issue is the interplay between optimisation and randomness, and the fact that attaining highly accurate estimations of the gradient at each step in \ac{SGD} can come at a high cost when dealing with large datasets. However, if we can accept certain levels of randomness in our gradient computation, then each step in the gradient descent can be achieved at a lower computational cost.
Returning to the context of biomedical optics, we may be able to accept a ``noisy'' low-cost forward model computation (which would otherwise be undesirable in the PDE-constrained approach) and simulate fewer photon trajectories during the earlier stages of the inversion process, leading to an overall accuracy vs computation time benefit. Thus, the topic of how to most 
efficiently utilize finite sized data sets in machine learning is relevant to the deployment of Monte Carlo based solvers in biomedical optics.

In this study we attempt to translate these recent insights from \ac{SGD} in machine learning into practical suggestions to improve the use of Monte Carlo methods in inverse problems which arise in biomedical optics.
To do this, we employ a fully stochastic computation of an objective gradient using forward and adjoint models of the RTE solved by the Monte Carlo method. This allows for the full non-linear inverse problem to be approached. In our demonstration problems the inverse problem can be approximately solved using a total computational expenditure which is similar or less than that which would typically be dedicated to a \emph{single} high quality (low variance) solution of the forward imaging problem. 

This paper is arranged as follows. 
First we will begin by outlining some key aspects of Gradient Descent in Section~\ref{sec:modelling}, including what appropriate metrics can be used to quantify acceptable levels of variance in the computation of sub-gradients via a stochastic process (i.e. Monte Carlo), and what step sizes to use in order to allow convergence.
In Section~\ref{sec:oracle} we will then describe the example problems that we will use to evaluate the improvements of \ac{SGD} in a biomedical context, as well as the details of the Monte Carlo forward and adjoint models and gradient calculations. 
In Section~\ref{sec:results} we apply these ideas to two different \emph{Coupled Physics Imaging} (CPI) modalities, namely \acf{QPAT} and \acf{UMOT}~\cite{AmmariBook2011,ArSc12}. 
Both of these problems are non-linear and entail the RTE for an accurate description, but exhibit different degrees of ill-posedness and resolution; thus they serve to demonstrate the generality of our approach. 
In Section~\ref{sec:results} we then evaluate the performance of various Monte Carlo inversions using simulated \ac{QPAT} and \ac{UMOT} data, and discuss what practical lessons can be taken from this in Section~\ref{sec:discussion}.

\section{Modelling and Inversion Problems in Optical Tomography}
\label{sec:modelling}

A common problem in biomedical optics involves finding the internal distribution of some optical properties $x$ within a medium using various measurements made around and/or within the medium, $\yobs$. To do this, we can employ some forward model of the underlying physical problem $A$, which produces an output $\ydata$, given some estimate of the internal properties $x$,
\begin{equation}
    \ydata = A\,x\;,
\end{equation}
where in this case the forward model $A$ could represent the radiative transport equation, and all relevant aspects of the optical setup (geometry of \Changes{sources} \& detectors etc.). In cases where $A$ is not directly invertible, then in order to solve for an unknown distribution of properties $x$, we can formulate a cost function as a measure of the quality of an estimate. This could for example be the L2-norm of the residual between the real measured data, $\yobs$, and our forward modelled data, $\ydata$:
\begin{equation}
    \Obj(x) = \frac{1}{2}\norm{\yobs - \ydata}^{2} = \frac{1}{2}\norm{\yobs - A\, x}^{2}
\end{equation}
From this point, the problem now becomes one of minimization, where we will qualify our solution $x^{\ast}$ as that which minimizes the cost function, $x^{\ast} = \argminA_{x} \Obj(x)$. Note that the ground truth parameters $x^{\rm true}$ may differ from the minimizer $x^{\ast}$ leading to \emph{reconstruction error}. 
This minimization problem can be approached via iterative Gradient Descent (GD), where we start with some estimate $x_{0}$, and each succesive iterate, $x_{n}$, is determined by subtracting a (scaled) gradient of our cost function $\nabla \Obj$ (relative to the internal optical properties) from the previous iterate,
\begin{equation}
\begin{split}
          x_{n} = x_{n-1} - \alpha_{n} \nabla \Obj(x_{n-1}) \;,
\end{split}
\label{eqn:GD}
\end{equation}
where $\alpha_{n}$ is the step size which scales the update term. 
If we have access to some computation or set of computations (sometimes referred to as a ``first-order oracle'') which we can call to compute $\Obj(x_{n-1})$ and $\nabla \Obj(x_{n-1})$, then this algorithm can be implemented, and is said to converge if $\lim_{n\to\infty} \Obj(x_{n}) = 0$. In practice, the descent may be terminated early once the cost function reaches some acceptable value, for example when the norm of the difference between observed and model data is of the same order as measurement noise, a criterion known as the \emph{Discrepancy Principle}~\cite{morozov66}.
\subsection{Stochastic Gradient Descent}
\label{sec:SGD}
In a stochastic setting, for instance when our forward model $A$ is a Monte Carlo model of radiative transport, then the true cost $\Obj$ and gradient $\nabla \Obj$ stated in \eref{eqn:GD} are not directly available. Instead, we may only have access to \textit{estimates} of the cost function and gradient (provided by a ``stochastic first-order oracle''). 
In Section~\ref{sec:oracle} we will detail the nature of these stochastic Monte Carlo computations in the radiative transport setting. In the interest of generality, for now we will simply assume such models exist, and that we can make a call to a ``stochastic oracle'' to attain $\Obj_{S_{n}}$ and $\nabla \Obj_{S_{n}}$ which we assume are non-biased approximations, i.e.
\begin{equation}
    \begin{split}
        \E[\Obj_{S_{n}}(x_{n})] & = \Obj(x_{n}) \;,\\
        \E[\nabla \Obj_{S_{n}}(x_{n})] & = \nabla \Obj(x_{n}) \;,
    \end{split}
\end{equation}
where $\E$ denotes the mean (expected) value for scalar quantities, or the mean (expected) vector for vector quantities such as the gradient.
Here $S_{n}$ denotes the $n$'th ``sample'' used in the computation. The meaning of ``sample'' here depends on the application. For example in machine learning this may refer to a particular training example (or group of training examples) to be used during one learning iteration~\cite{johnson2013}. 
In Monte Carlo modelling of radiative transport, the sample refers to the set of 
virtual photons (and their associated random number seeds) that are initiated in 
the simulation to represent an optical source, which are subsequently used to 
estimate $\Obj(x_{n})$ and $\nabla \Obj(x_{n})$. The stochastic version of 
Gradient Descent (\ac{SGD}) thus attempts to minimize a \textit{sampled} 
objective function, $\Obj_{S_{n}}$, by updating the previous iterate with a scaled \textit{sampled} gradient, 
\begin{equation}
\begin{split}
          x_{n} = x_{n-1} - \alpha_{n} \nabla \Obj_{S_{n}}(x_{n-1}) \;.
\end{split}
\label{eqn:SGD}
\end{equation}
As with any computation, a call to a stochastic oracle at each iteration comes with a certain computational cost. The particular cost may depend on a number of factors, including the sample size, $\vert S_{n} \vert$. This 
is one of the reasons \Changes{why} the study of \ac{SGD} is of such importance in modern machine learning, where training data sets may be of an enormous 
size, meaning that computing a gradient based on all available data at each iteration could be very costly. Rather, individual samples ($\vert S_{n} \vert = 1$), or batches of samples ($\vert S_{n} \vert > 1$) may be 
used instead at each iteration. While this degrades the quality of any individual gradient estimate compared to using all available data, if the 
variance of these estimates is maintained below an acceptable value, the overall tradeoff may be net positive.
What this means in a Monte Carlo radiative transport context is that we may be able to allow low quality gradient estimations (simulating only a 
small number photons) for a large part of the inversion process when estimating optical properties, saving on per iteration computational 
resources, leading to an overall efficiency improvement. This is in contrast to typical implementations of iterative Monte Carlo solvers in 
the biomedical optics community, where each iteration is computed with large numbers of photons that are deemed sufficient to produce ``stable'' 
and ``smooth'' (low variance) forward model data~\cite{friebel2006determination,yaroslavsky1996inverse,fang2009,fang2010,buchmann2019three,hochuli2016quantitative,leino2019}. 
In some cases, where a linearised approximation is assumed for the inverse problem, the cost of rerunning the forward model can be avoided using techniques such as Perturbation Monte Carlo (PMC) methods~\cite{hayakawa2001perturbation,tricoli2017reciprocity,yao2018}. However, for the full non-linear problem, although PMC can be used for calculation of the problem Jacobian, this has to be recomputed at each iteration of, for example, a Gauss-Newton optimisation scheme~\cite{leino2019perturbation}.

If in this study we are to accept a level of variance and imperfection in our forward/adjoint models, this of course raises the question of how much variance is acceptable in order for \ac{SGD} to be successful? Furthermore, what measure of the variance is the best indicator in terms of efficiency/performance for common Monte Carlo solvers? To begin to answer this, it is important to first note that fixed-step \ac{SGD} does not in general converge to a solution\cite{bottou2018optimization,newton2018stochastic}. That is, if $\alpha_{n}$ is fixed for all $n$, eventually there will come a point where the next update of the estimate (with the term $\alpha_{n} \nabla \Obj_{S_{n}}(x_{n-1})$) will reliably ``undo'' the work of the prior step, which will effectively halt the descent. The point at which this occurs depends on the variance of $\nabla \Obj_{S_{n}}$. We can see this by re-writing the sampled stochastic gradient estimate as,
\begin{equation}
    \nabla \Obj_{S_{n}}(x_{n}) = \nabla \Obj(x_{n}) + \epsilon_{S_{n}}(x_{n}) \; ,
\end{equation}
where $\epsilon$ is a random vector with $\E[\epsilon_{S_{n}}(x_{n})] = 0$ for all $n$. As gradient descent progresses successfully, the ``true'' gradient $\nabla \Obj$ will eventually begin reducing in size as we near the minimum. Once the magnitude of the true gradient reduces to a point at which it is comparable to the randomness of $\epsilon_{S_{n}}$, the problem arises. The larger the expected magnitude of $\epsilon_{S_{n}}$, the sooner the minimization of the cost function reaches this limiting scenario, where further iterations will only lead to a random walk about this point.

To prevent this from happening, we may take one of two actions (or a combination thereof): i) reduce the step size at each iteration such that we can avoid ``backtracking'' in the descent, more on this in Section~\ref{sec:stepsize}; or ii) gradually improve the accuracy of our sampled gradient such that the variance of the sampled gradient remains below some threshold value compared to the norm of the true gradient $\nabla \Obj$.
In other words, we may wish to ensure the inequality
\begin{equation}
V^{2}_{\rm tot}(x_{n}) :=    \frac{\E \left [ \norm{ \epsilon_{S_{n}}(x_{n})}^{2}\right ]}{\norm{\nabla \Obj(x_{n})}^{2}}  \leq \gamma_{\rm tot}^{2} \;,\;\;\;\; \gamma_{\rm tot} > 0.
\label{eqn:norm}
\end{equation}
where $\gamma_{\rm tot}$ is a positive coefficient describing the acceptable threshold. 
The above inequality is known as the ``norm test''\cite{bolla2018adaptive}. Note that, since for any vector of random variables the variance of its length is the sum of the variances parallel and orthogonal to any fixed vector, this test equally penalizes the components of randomness parallel and perpendicular to the true gradient.
Recent studies however have demonstrated that controlling the component of randomness parallel to $\nabla \Obj$ is potentially a more relevant objective, as the component of the sampled gradient orthogonal to the true gradient is zero in expectation. An alternative measure of acceptable variance in $\nabla \Obj_{S_{n}}$ has thus been introduced as the ``inner product test''~\cite{bolla2018adaptive}, which only aims to restrict the component of variance in the sampled gradient parallel to the true gradient $\nabla \Obj$,
\begin{equation}
\Changes{
V^{2}_{\parallel}(x_{n}) :=    \frac{\E \left [ \left < \epsilon_{S_{n}}(x_{n}),\nabla \Obj(x_{n}) \right > ^{2} \right ]}{\norm{\nabla \Obj(x_{n})}^{4}} \leq \gamma_{\parallel}^{2}  \;,\;\;\;\; \gamma_{\parallel} > 0.}
\label{eqn:inner}
\end{equation}
This inner product test imposes a less restrictive limitation of the overall variance in the sampled gradients, particularly in cases where the variance may be higher in directions orthogonal to the true gradient than in the direction parallel to $\nabla \Obj$.
Either of these metrics however will be able to exploit the fact that an increased 
expected error, $\E \left [ \norm{\epsilon_{S_{n}}} \right ]$, will correlate to a cheaper 
computation of the estimated gradient. Thus, setting larger values of $\gamma_{\rm tot}$ or $\gamma_{\parallel}$ in the inequalities will correspond to cheaper computational 
requirements for each step, but also a more pronounced random walk component to the 
gradient descent. In many cases, it may be found that the penalty paid by 
increasing the random walk component is acceptable (up to a point) compared to the 
penalty paid in computational cost for reducing the expected norm of $\epsilon$ to 
a negligible value. For example, using Monte Carlo RTE simulations to compute $\nabla \Obj$ with a negligible level of variance (i.e. setting $\gamma_{\rm tot} \ll 1 $) may take billions of simulated photons at each step. Whereas, it may be 
possible to compute a gradient that passes the norm test or inner product with larger values of $\gamma_{\rm tot}$ or $\gamma_{\parallel}$ with many orders of magnitude less photons, particularly 
during the early stages of gradient descent, where we may be far from the minimum. The 
ideal choice of $\gamma_{\rm tot}$, or $\gamma_{\parallel}$ will depend on the 
specific application.

\subsection{Adaptive sample size}
\label{sec:sampsize}
We have discussed two different measures of the variance in the sampled gradient $\nabla \Obj_{S_{n}}$ that we wish to investigate in the context of Monte Carlo estimation of media properties, \emph{viz.} the norm test \eref{eqn:norm}, and the inner product test \eref{eqn:inner}. 
In order to satisfy the inequalities defining these tests as the gradient descent progresses, we will be required to reduce the variance in the sampled gradients $\nabla \Obj_{S_{n}}$ whenever the norm test or inner product test fail. This can be done by increasing the sample size (number of photons used, $\vert S_{n} \vert$)  when making a call to the stochastic oracle. 
Two practical considerations are still required: first, how to compute the ``true'' gradient $\nabla \Obj$, which is needed to evaluate the norm test and inner product test; and second, by how much we should increase the sample size in a situation where one of the tests fails?

The true gradient $\nabla \Obj$ is only calculable in the limit that an infinite number of photons are used in the Monte Carlo model. \Changes{This limit can equivalently be represented as an average over independent repeated outputs of the sampled gradient,
\begin{equation}
    \nabla \Obj (x_{n})  = \lim_{N_{\rm rep} \to \infty } \frac{1}{N_{\rm rep}}\sum_{j = 1}^{N_{\rm rep}} \nabla \Obj_{S_{j}}(x_{n})\;\;\;,\;\;\;\; \vert S_{j} \vert = \vert S_{n} \vert \;\; \forall \;\;j \;\;.
\label{eqn:av_grad}
\end{equation}
Using a finite value of $N_{\rm rep}$ in the evaluation of \eref{eqn:av_grad} provides an approximation to the true gradient, and when this is used to compute the norm and inner product tests (Eqs.~\ref{eqn:norm} \& \ref{eqn:inner}) the inequalities will fail \textit{before} they would if $N_{\rm rep} = \infty$, thus acting as a conservative approximation.} 
It is noted that this method of approximating the true gradient is computationally taxing. In practice however, the inner product test and norm tests can still be conducted efficiently if they are only computed occasionally (not at every iteration) of the descent. For example, using $N_{\rm rep} = 100$ repeated computations of the sampled gradient to conduct the tests once every 100 iterations (thus only updating our sample size every 100 iterations) would only double the total number of simulated photons required for the inversion. In this study we evaluate these metrics once every 10 iterations, using $N_{\rm rep} = 100$ repeated sampled gradients. While this is a significant computational burden, we do so in this study as we are interested in assessing the best case scenarios for such methods. Note that although we compute the above approximation to the true gradient for the purposes of evaluating the inner product and norm tests, we only ever update our estimate using the sampled gradient. 

In terms of increasing the sample size in the event where the inner product and/or norm tests fail, this can be done in a number of ways. A simple method we will employ in this study is to scale the current sample size by some factor $\kappa(n)$, to increase the number of photons used in the next iteration, 
\begin{equation}
    \vert S_{n+1} \vert = \kappa(n)\,\vert S_{n} \vert
\end{equation}
One option for $\kappa(n)$ is to use the same factor by which the variance 
exceeds our imposed limit at a given point in the descent. For instance, upon failure of the inner product test for a chosen value of $\gamma_{\parallel}$, we can increase the sample size on the next iteration 
using $\kappa(n) = V^{2}_{\parallel}(x_{n})/\gamma_{\parallel}^{2}$. However, we also investigate other forms of $\kappa(n)$ in the Section~\ref{sec:results}, which better cope with statistical variations that can lead to over-estimating the required sample size increase.

\subsection{Step size}
\label{sec:stepsize}
In cases where we are not taking actions to bound the error in the sampled gradient (such as enforcing successful outcomes of an inner product test or norm test), fixed step \ac{SGD} may only converge to a region around the solution. Reducing the step size sufficiently at each step is usually required to allow convergence~\cite{robbins1951stochastic}. However, it can be shown that if we are bounding the error in the sampled gradient, e.g. by increasing the sample size, then fixed step \ac{SGD} may converge so long as the following is satisfied for all $n$~\cite{bolla2018adaptive} 
\begin{equation}
    \alpha_{n} \leq \frac{1}{(1 + \gamma_{\rm tot}^{2})L} \; \;,
\label{eqn:step}
\end{equation}
where $L$ is the Lipschitz constant for $F$\footnote{The Lipschitz constant for a functional $F$ is a measure of its rate of change with respect to its parameter and can be defined for example as the smallest constant such that $\nabla^2 \Obj \preceq L\,\rm{Id}$, where $\rm{Id}$ is the identity matrix, and we assume that $\Obj$ is twice continuously differentiable. It can also be interpreted as the largest eigenvalue of the Hessian of $\Obj$~\cite{nesterov2013introductory}.}. 
As intuition may indicate, when the sample size (e.g number of simulated photons) increases towards the maximum number of samples $\vert S_{n} \vert \rightarrow \vert S_{\rm max} \vert$ ($\vert S_{\rm max} \vert = \infty$ in the case of Monte Carlo RTE simulations), the expected error in the sampled gradient approaches zero, $\vert \epsilon_{S_{n}} \vert \rightarrow 0$, as do the measures of variance in the sampled gradients ($V^{2}_{\rm tot} \rightarrow 0$, $V^{2}_{\parallel} \rightarrow 0$), 
as defined in \eref{eqn:norm} and \eref{eqn:inner}. 
In other words, as the stochasticity in the problem reduces to zero, 
we approach the classical step size 
of the deterministic problem given by $\alpha = \frac{1}{L}$~\cite{nesterov2013introductory}.

In this study we will aim to satisfy the above step size criteria for an assumed 
value of the Lipschitz constant $L$, which we will choose conservatively depending 
on the particular scenario. However, as we are primarily interested in reaching 
the best possible solution for a given allocation of computational resources, 
convergence to a region around the unique solution may in fact be sufficient for 
our purposes. For this reason, we will also investigate larger step size criteria 
which violate \eref{eqn:step}, yet exhibit good performance in our scenarios of 
interest.

With these considerations in mind, below we present in Algorithm~\ref{alg:adaptive} a basic method for Stochastic Gradient Descent using adaptive sample sizes (simplified from Ref.~\cite{bolla2018adaptive}). The algorithm imposes a limit on the total number of photons to be simulated using Monte Carlo transport models throughout the entire descent, $N_{\rm ph}$. 
\begin{algorithm}
\begin{algorithmic}
\STATE Choose initial photon sample size $\vert S_{1} \vert$, and desired value of $\gamma_{\parallel}$, or $\gamma_{\rm tot}$
\WHILE{$\sum_{i=1}^{n} \vert S_{i}\vert < N_{\rm ph}$}
\IF{run test?}
\STATE{compute sampled gradient, $\nabla \Obj_{S_{n}}$, and \Changes{\textit{approximate}} true gradient, $\nabla \Obj$ \Changes{(using \eref{eqn:av_grad})}}
\STATE{check norm test (or) inner product test is satisfied}
\IF{test fail} \STATE{increase sample size on next iteration $\vert S_{n+1} \vert =\kappa(n) \, \vert S_{n} \vert$} 
\ELSE \STATE{set $\vert S_{n+1} \vert =\vert S_{n} \vert$}
\ENDIF 
\ELSE 
\STATE{compute sampled gradient only $\nabla \Obj_{S_{n}}$}
\STATE{set $\vert S_{n+1} \vert =\vert S_{n} \vert$}
\ENDIF

\STATE update $x_{n+1} = x_{n} - \alpha_{n}\nabla \Obj_{S_{n}}$

\ENDWHILE
\end{algorithmic}
\caption{Inversion using Monte Carlo sampled gradients with adaptive sample size}
\label{alg:adaptive}
\end{algorithm}

\section{Stochastic forward and adjoint models}
\label{sec:oracle}

In this section we cover the computation of the stochastic forward model and stochastic gradient approximation, referred to above as the first-order stochastic oracle. We will cover the basic radiative transport forward problem, as well as the gradient computations involved in our example problems of absorption estimation in \Changes{\acf{QPAT} and \acf{UMOT}}. The specific details of these models are not required to understand the main premise of this paper, but serve as a demonstration in a context familiar to many in the biomedical optics community, where Monte Carlo models of optical transport are employed to estimate medium properties. 
\subsection{Forward model}
For any optical source $Q(\vec{r},\hat{\vec{s}})$, either incident on a medium or present  within it, we wish to model the resulting radiance, $\phi(\vec{r},\hat{\vec{s}})$, 
describing the radiant flux at each position $\vec{r}$, and in each direction $\hat{\vec{s}}$. This can be achieved using the Radiative Transport Equation (RTE):
\begin{equation}
   \underbrace{\left( \hat{\vec{s}} \cdot \nabla + \mua(\vec{r}) + \mus(\vec{r}) \right)}_{\mathcal{T}_{\mua,\mus}} \phi(\vec{r}, \hat{\vec{s}}) = \underbrace{\mus(\vec{r})\int_{\mathcal{S}^{2}} p(\hat{\vec{s}}, \hat{\vec{s}}')}_{\mathcal{S}_{\mus}} \phi(\vec{r}, \hat{\vec{s}}') \; \textrm{d}\hat{\vec{s}} + Q(\vec{r},\hat{\vec{s}})\;.
    \label{eq:RTE1}
\end{equation}
where we denote $\mathcal{T}$, and $\mathcal{S}$, as the attenuation and scattering operators, which together compose the RTE operator, $\mathcal{L}$ \footnote{For notational convenience we assume that \eref{eq:RTE1} is combined with appropriate boundary conditions which we do not write explicitly here; see~\cite{arridge99} for more details.}. \Changes{Here, $\mu_{a}$ is the absorption coefficient, $\mu_{s}$ is the scattering coefficient, and $p$ is the scattering phase function. Using the defined operators, \eref{eq:RTE1} can be rewritten in a more compact form}
\begin{equation}
    \mathcal{L}_{\mu_{\rm a},\mu_{\rm s}}\phi = \left ( \mathcal{T}_{\mu_{\rm a},\mu_{\rm s}}- \mathcal{S}_{\mu_{\rm s}} \right )\phi = Q\;.
\end{equation}
In order to obtain (stochastic) estimates of the radiance resulting from a given 
source, and thus to obtain an estimate of any derived data function $y(\phi)$, 
we can implement a Monte Carlo solver, $\MC$.
In this study we have adapted a GPU-accelerated version of the commonly employed ``Monte Carlo Multi-Layer'' (MCML) program used to simulate radiative transport within a layered planar medium~\cite{erik2009cudamcml,wang1995mcml}. The basic operation of this program is unchanged from the original release. Simulated photons are initiated by sampling from a given source function, $Q$, and scattering/absorption events are pseudo-randomly generated along each photon's trajectory until either: i) the photon leaves the domain, or; ii) the photon drops its weight below some threshold value. The scattering directions are sampled from the Henyey-Greenstein scattering phase function in this study.
The expected accuracy of the computed radiance using such Monte Carlo solvers $\MC$ depends on the total number of photons used, i.e. the sample size $\vert S_{n} \vert$. As $\vert S_{n} \vert \rightarrow \infty$, the radiance approaches the deterministic solution of the RTE. Importantly however, 
Monte Carlo models allow an estimate of the radiance to be achieved with \textit{any} number of photons with $\vert S_{n} \vert \geq 1$. 
The expected computational requirements (number of floating point operations) of the Monte Carlo solver $\MC$ also scales with the number of photons simulated, and it is this trade-off between accuracy of the forward model (and corresponding adjoint model) and computational cost that we will be investigating. 

\subsection{Gradient computation: Adjoint model}
To compute the gradient of our cost function $\nabla \Obj$, with respect to the optical properties of the medium, we make use of an adjoint \ac{RTE} model. 
\Changes{Although direct methods of finding the derivative of a Monte Carlo method can also be developed~\cite{Graff93}, adjoint methods have more applicability in general, and also allow closer comparison with optimisation techniques used in machine learning. For further details of forward and adjoint methods in the \ac{RTE} we refer to~\cite{Bal2009}; for specific details of coupled physics imaging probems we refer to~\cite{Bal2013}.}
We first consider a change to \eref{eq:RTE1} 
where $\mua\rightarrow \mua + \mua^{\delta}, \mus\rightarrow \mus + \mus^{\delta}$, 
for the same source $Q$, which results in a change in radiance $\phi\rightarrow \phi+\phi^{\delta}$. This implies
\begin{eqnarray}
    \left(\mathcal{T}_{\mua+\mua^\delta,\mus+\mus^\delta} - \mathcal{S}_{\mus+\mus^\delta}\right)\left(\phi+\phi^\delta\right) &=& \left(\mathcal{T}_{\mua,\mus} - \mathcal{S}_{\mus}\right)\phi \nonumber \\
    \Rightarrow \quad \left(\mathcal{T}_{\mua,\mus} - \mathcal{S}_{\mus}\right)\phi^\delta &=& -(\mua^\delta+\mus^\delta + \mathcal{S}_{\mus^\delta})\phi \\
\mathcal{L}_{\mua,\mus} \phi^{\delta} &=& -\underbrace{
(\mua^\delta+\mus^\delta + \mathcal{S}_{\mus^{\delta}})
}_{
\mathcal{L}^{\delta}_{\mua^{\delta},\mus^{\delta}}}\phi\;.
    \label{eq:FrechetRTE}
\end{eqnarray}
We also define the fluence, $\Phi$, as the angular integral of the radiance,
\begin{equation}
    \Phi(\vec{r}) = \int_{\mathcal{S}^{2}}\phi(\vec{r},\hat{\vec{s}})\,d\hat{\vec{s}}\;.  
\label{eqn:fluence}
\end{equation}
To proceed beyond this point, we must now consider the specific form of the data function relevant to a particular modality of interest. We begin with the first of our two example modalities, \acf{QPAT}.

\subsubsection{\ac{QPAT} case\label{sect:QPAT}}
\begin{figure}[h]
\centering
\includegraphics[width=7cm]{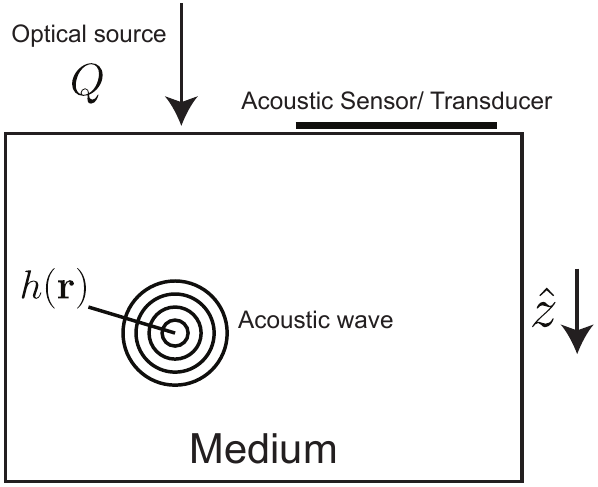}
\caption{\small{Setup for Quantitative Photoacoustic Tomography.}}
\label{fig:QPAT_setup}
\end{figure}

In \ac{QPAT} the medium is illuminated with a pulsed optical source, Q \Changes{(see Fig.~\ref{fig:QPAT_setup})}. The distributed optical energy is absorbed at various points within the sample, giving rise to internal acoustic waves. These acoustic waves can be detected at the surface of the medium by a sensor, and processed to locate the initial pressure distribution $p_{0}$ within the medium~\cite{Wa09,Bea11,NiCh14}. This internal pressure distribution is related to the spatial distribution of absorbed optical energy, $\qdata$, where 
\begin{equation}
    \qdata(\vec{r}) = \mua(\vec{r}) \Phi(\vec{r}),
\end{equation}
and where $\Phi$ is the optical fluence of \eref{eqn:fluence} \footnote{We have omitted the Gr\"uneisen parameter for clarity of exposition, though this parameter can be included in practice.}. Assuming that we can recover the absorbed optical energy, $\qdata$, the problem remains to find the distribution 
of $\mu_{\rm a}(\vec{r})$ within the medium\cite{cox2012,saratoon2013}. 
Note that although the optical source is pulsed, it is acceptable to use a continuous wave (time-independent) model to describe $\phi$ and $\Phi$ because the time scale of the acoustic wave propagation is orders of magnitude slower than the optical propagation~\cite{WaAn11}. 
First, restating our cost function in terms of the \ac{QPAT} data function, $h$, we have
\begin{equation}
    \qObj \quad = \quad\Changes{\frac{1}{2}}\int_{\domain}(\qobs-\qdata)^2 \rmd \pos \quad =\quad  \frac{1}{2}\left<\qobs-\qdata,\qobs-\qdata \right>_{L^2(\domain)}\;.
\end{equation}
We then write the \Frechet \,derivative of $\qObj$ as 
\begin{equation}
D\qObj = -\left<\qobs-\qdata,D\qdata \mua^{\delta}\right>_{L^2(\domain)}\;,
\end{equation}
where $\mua^\delta$ is a small change in absorption. In this paper we will neglect changes in scattering\Changes{, however the below formalism is still general for the gradient with respect to absorption. The gradient term with respect to scattering coefficient is described for example in~\cite{saratoon2013}, and will be included in future investigations}. Writing the \Frechet \,derivative of $\qdata$ as
\begin{equation}
    D\qdata = \Phi + \mua\cdot D\Phi\;,
\end{equation}
and defining $\Phi^\delta = D\Phi \mua^\delta$, we arrive at 
\begin{equation}
D\qObj = -\left<\Phi(\qobs-\qdata),\mua^{\delta}\right>_{L^2(\domain)} - \left<\mua(\qobs-\qdata),\Phi^{\delta}\right>_{L^2(\domain)}\;.
\label{eq:QPAT_RTE_2}
\end{equation}
Next, we define the adjoint radiance, $\phi^{\ast}$, as the solution to
\begin{equation}
    \mathcal{L}^{\ast}\phi^{\ast} = \mua(\qobs-\qdata)
    \label{eq:QPATadjrad1}
\end{equation}
where the right hand side describes the ``adjoint source'' which is isotropic in $\hat{\vec{s}}$. We then substitute the above into \eref{eq:QPAT_RTE_2} to give 
\begin{equation}
D\qObj = -\left<\Phi(\qobs-\qdata),\mua^{\delta}\right>_{L^2(\domain)} - \left<\mathcal{L}^{\ast}\phi^{\ast},\phi^{\delta}\right>_{L^2(\domain\times S^{n-1})}
\label{eq:QPAT_RTE_3}
\end{equation}
where we exploited the fact that the right hand side of \eref{eq:QPATadjrad1}
does not depend on direction. Using the definition of the adjoint operator, and the fact that the change in radiance is zero on the boundary $\bdomain$ yields
\begin{equation}
D\qObj = -\left<\Phi(\qobs-\qdata),\mua^{\delta}\right>_{L^2(\domain)} - \left<\phi^{\ast},\mathcal{L}\phi^{\delta}\right>_{L^2(\domain\times S^{n-1})}\,.
\label{eq:QPAT_RTE_4}
\end{equation}
Finally we make use of the perturbation expression \eref{eq:FrechetRTE}, whilst again here we neglect any change in scattering. This gives
\begin{equation}
D\qObj = -\left<\Phi(\qobs-\qdata),\mua^{\delta}\right>_{L^2(\domain)} + \left<\phi^{\ast}\phi,\mua^{\delta}\right>_{L^2(\domain\times S^{n-1})}\,,
\label{eq:QPAT_RTE_5}
\end{equation}
allowing us to define the (absorption) gradient as in Eq. (33) of~\cite{saratoon2013} :
\begin{equation}
    \frac{\partial\qObj}{\partial\mua} = \nabla \qObj =  -\Phi(\qobs-\qdata) + \int_{S^{n-1}}\hspace{-1em}\phi^{\ast}\phi\, \rmd\angvec
\label{eqn:grad_QPAT}
\end{equation}
To compute a stochastic approximation of this gradient, we can thus use the 
forward model Monte Carlo solver $\MC$ to provide estimates of $\phi$ and $\Phi$, and an adjoint Monte Carlo solver $\AMC$ to produce $\phi^{*}$ from an adjoint source term $ Q_{\textup{adj}} = \mu_{\rm a}(\qobs-\qdata)$, as defined in \eref{eq:QPATadjrad1}. Due to the symmetry of the problem, the adjoint solver is identical to the forward solver, and follows the same basic operating principles. The only difference is that here the adjoint source $ Q_{\rm adj} = \mu_{\rm a}(\qobs-\qdata)$ may in fact be negative in some locations. This is handled by splitting the source term into two parts, one purely positive, $Q_{\rm adj}^{+}$, and one purely negative, $Q_{\rm adj}^{-}$. Two simulations are then run (where the total number of photons to be used is split between the two simulations accordingly), and the results summed to produce $\phi^{*}$.
The following algorithm describes the basic operation for computing a sampled gradient, $\nabla \Obj_{S_{n}}$, for \ac{QPAT} using the above derivation. This will be used in conjunction with Algorithm~\ref{alg:adaptive} to conduct an inversion with adaptive sample size for each iterate, $\vert S_{n}\vert$.

\begin{algorithm}
\begin{algorithmic}

\STATE 1) Compute $\MC Q\mapsto \phi,\,\Phi$, using $\vert S_{n} \vert / 2$ photons 
\STATE 2) Construct internal adjoint source $Q_{\textup{adj}} = \mua(\qobs-\qdata)$
\STATE 3) Compute $\AMC Q_{\textup{adj}}\mapsto \phi^{*},\,\Phi^{*}$, using $\vert S_{n} \vert / 2$ photons 
\STATE 4) Use \eref{eqn:grad_QPAT} to compute gradient $\nabla \Obj_{S_{n}}$
\end{algorithmic}
\caption{Monte Carlo sampled \ac{QPAT} gradient}
\label{alg:QPAT}
\end{algorithm}

\subsubsection{\ac{UMOT} case}
\label{sec:UMOT_case}

\begin{figure}[h]
\centering
\includegraphics[width=7.0cm]{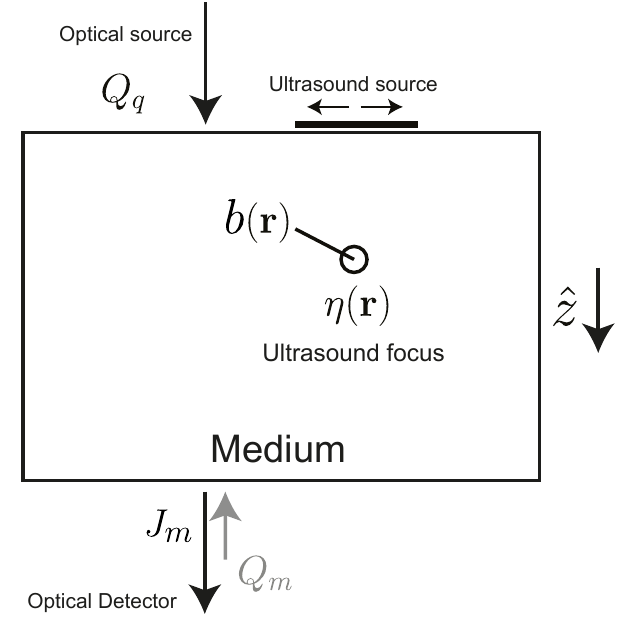}
\caption{\small{Setup for Ultrasound-Modulated Optical Tomography in the transmission geometry.}}
\label{fig:UMOT_setup}
\end{figure}

\Changes{Referring to Figure~\ref{fig:UMOT_setup}}, in \acl{UMOT} we have an optical light 
source $Q_{q}$ incident on a medium, as well as an optical detector $J_{m}$. In addition, an ultrasound source is incident on the medium, where the focus $\eta(\vec{r})$ is scanned through the sample~\cite{Ammari2014ARA,Chung2017InverseTA}.
Assuming for simplicity an ideal (delta-function) ultrasound focus, the data of interest in this case is found to be of the form~\cite{powell2015gradient} 
\begin{equation}
    \udata(r) = \eta(\vec{r}) \Phi_{q}(\vec{r}) \Phi_{m}(\vec{r}),
    \label{eqn:UMOTdata}
\end{equation}
where $\Phi_{q}$ is the fluence resulting from the optical source $Q_{q}$,  and $\Phi_{m}$ is the resulting fluence from a virtual source $Q_{m}$ which is reciprocal to the detector $J_{m}$~\cite{powell2015gradient}. 
From this point we proceed in similar fashion as in section~\ref{sect:QPAT}, where now our data fitting error is given by
\begin{equation}
    \uObj \quad = \quad\Changes{\frac{1}{2}}\int_{\domain}(\uobs-\udata)^2 \rmd \pos \quad =\quad  \frac{1}{2}\left<\uobs-\udata,\uobs-\udata \right>_{L^2(\domain)}
\end{equation}
and its \Frechet \,derivative as 
\begin{equation}
D\uObj = -\left<\uobs-\udata,D\udata \mua^{\delta}\right>_{L^2(\domain)}
\end{equation}
In this case the \Frechet \,derivative of $\udata$ becomes
\begin{equation}
    D\udata = \eta\Phi_{q}\cdot D\Phi_{m} + \eta\Phi_{m}\cdot D\Phi_{q}
\end{equation}
leading to  
\begin{equation}
D\uObj = -\left<\eta\Phi_{q}(\uobs-\udata),\Phi_m^{\delta}\right>_{L^2(\domain)} - \left<\eta\Phi_{m}(\qobs-\qdata),\Phi_q^{\delta}\right>_{L^2(\domain)}\;.
\label{eq:UMOT_RTE_2}
\end{equation}
Here we need to define two adjoint radiances, $\phi^{\ast,1}$, $\phi^{\ast,2}$, as the solution to
\begin{eqnarray}
    \mathcal{L}^{\ast}\phi^{\ast,1} &=&\eta\Phi_{q}(\uobs-\udata)
    \label{eq:UMOTadjrad1}\\
        \mathcal{L}^{\ast}\phi^{\ast,2} &=&\eta\Phi_{m}(\uobs-\udata)
    \label{eq:UMOTadjrad2}
\end{eqnarray}
and substituting into \eref{eq:UMOT_RTE_2} to give 
\begin{equation}
D\uObj =  - \left<\mathcal{L}^{\ast}\phi^{\ast,1},\phi_m^{\delta}\right>_{L^2(\domain\times S^{n-1})} - \left<\mathcal{L}^{\ast}\phi^{\ast,2},\phi_q^{\delta}\right>_{L^2(\domain\times S^{n-1})}
\label{eq:UMOT_RTE_3}
\end{equation}
by the same arguments as for \ac{QPAT} we get 
\begin{equation}
D\uObj =  - \left<\phi^{\ast,1},\mathcal{L}\phi_m^{\delta}\right>_{L^2(\domain\times S^{n-1})} - \left<\phi^{\ast,2},\mathcal{L}\phi_q^{\delta}\right>_{L^2(\domain\times S^{n-1})}\,.
\label{eq:UMOT_RTE_4}
\end{equation}
Again using the perturbation expression \eref{eq:FrechetRTE} we have
\begin{equation}
D\uObj =  \left<\phi^{\ast,1}\phi_m,\mua^{\delta}\right>_{L^2(\domain\times S^{n-1})}+ \left<\phi^{\ast,2}\phi_q,\mua^{\delta}\right>_{L^2(\domain\times S^{n-1})}\,.
\label{eq:UMOT_RTE_5}
\end{equation}
allowing us to define the (absorption) gradient as 
\begin{equation}
    \frac{\partial\uObj}{\partial\mua} = \nabla \uObj =   \int_{S^{n-1}}\left(\phi^{\ast,1}\phi_m + \phi^{\ast,2}\phi_q\right)\, \rmd\angvec
\label{eqn:grad_UMOT}
\end{equation}
Thus, similar to the \ac{QPAT} case, here we are able to compute a stochastic approximation of this gradient using the forward model Monte Carlo solver $\MC$ to provide $\phi_{q}$ and $\phi_{m}$ from our two sources, and an adjoint Monte Carlo solver $\AMC$ to produce $\phi^{*\,,1}$ and $\phi^{*\,,2}$ from the adjoint source terms $Q_{\textup{adj}}^{1} = \eta \phi_{q}(\uobs-\udata)$ and $Q_{\textup{adj}}^{2} = \eta \phi_{m}(\uobs-\udata)$, as defined in Eqs.~(\ref{eq:UMOTadjrad1}--
\ref{eq:UMOTadjrad2}). Here as well, adjoint source terms are split into two parts, one purely positive, $Q_{\rm adj}^{+}$, and one purely negative, $Q_{\rm adj}^{-}$, with the photon budget being split accordingly. 
The following algorithm describes the basic operation for computing a sampled gradient, $\nabla \Obj_{S_{n}}$, for \ac{UMOT} using the above derivation. This will be used in conjunction with Algorithm~\ref{alg:adaptive} to conduct an inversion with adaptive sample size for each iterate, $\vert S_{n}\vert$.
\begin{algorithm}
\begin{algorithmic}
\STATE 1) Compute $\MC Q_{q}\mapsto \phi_{q},\,\Phi_{q}$, and $\MC Q_{m}\mapsto \phi_{m},\,\Phi_{m}$, each using $\vert S_{n} \vert / 4$ photons 
\STATE 2) Construct internal adjoint sources $Q_{\textup{adj}}^{1} =  \eta\Phi_{q}(\uobs-\udata)$ and $Q_{\textup{adj}}^{2} =  \eta\Phi_{m}(\uobs-\udata)$
\STATE 3) Compute $\AMC Q_{\textup{adj}}^{1} \mapsto \phi^{*\,,1},\,\Phi^{*\,,1}$, and $\AMC Q_{\textup{adj}}^{1} \mapsto \phi^{*\,,2},\,\Phi^{*\,,2}$, each using $\vert S_{n} \vert / 4$ photons 
\STATE 4) Use \eref{eqn:grad_UMOT} to compute $\nabla \Obj_{S_{n}}$

\end{algorithmic}
\caption{Monte Carlo sampled \Changes{\ac{UMOT}} gradient}
\label{alg:UMOT}
\end{algorithm}

\subsection{Fluence Monte Carlo}

It should be noted that numerous Monte Carlo radiative transport solvers do not explicitly output the radiance, as this requires additional programming to store the angular ordinates at each location. Commonly, only the fluence will be 
available, which is the angular integral of the radiance \eref{eqn:fluence}. 
In such cases, the above integrals for the gradients of interest Eqs.~(\ref{eqn:grad_QPAT}),
(\ref{eqn:grad_UMOT}) can be computed under the assumption of approximately angularly isotropic radiances, where for example $\int \phi^{*} \phi\, d\hat{\vec{s}}$ becomes $\Phi^{*} \Phi$. The accuracy of this approximation of course depends on the true angular dependence of the radiances, where the approximation is poorest in regions close to directional light sources, but improves further away. The higher the scattering asymmetry $g$ of the medium, the slower the approximation improves as a function of distance from these sources. In many cases however this is a satisfactory assumption, and is employed in the below example cases.

\section{Results}
\label{sec:results}

In this section we present the results of a number of investigations using our 
two example problems of \ac{QPAT} and \ac{UMOT}. We will demonstrate the 
implementation of the forward-adjoint Monte Carlo solvers described above, along 
with adaptive sampling strategies to estimate the absorption coefficient of a 
medium via \ac{SGD}. \Changes{Here we investigate media with a semi-infinite slab 
geometry, with numerous layers in the z-direction having different optical 
properties, but otherwise homogeneous in the $x$ and $y$ directions. The application to layered geometry in this demonstration was chosen for simplicity to provide an easily recognizable setting to test these adaptive sampling methods.
Furthermore, whilst apparently simplistic, layered geometries are still of practical interest for applications including instrument calibration and validation, and the imaging of biological structures with small curvature but significant heterogeneity in depth. The latter example includes studies such as functional (cognitive) imaging when localised to small activation regions. Application of these new methods in more complicated 3D geometries will be carried out in future work.} Each of the medium layers can be described in terms of thickness, scattering coefficient, 
absorption coefficient, refractive index (background) and scattering asymmetry 
parameter. We will assume all parameters of the layered medium are known 
\textit{a priori} with the exception of the absorption coefficient, which we will 
attempt to solve for. For the examples in this study we set the total slab 
thickness to 2cm, and the inversion is conducted with a resolution of $0.25$mm,
(80 layers). The true ``measured'' data in all problems is generated using a 
single forward model Monte Carlo simulation using a large sample size of $10^{9}$ photons. With this sample size, the variance of the measured forward data $\qobs$, $\uobs$ is found to be negligible in this setup, and as such can be treated as effectively equivalent to the deterministic solution of the RTE.

To conduct an inversion, we stipulate a total photon budget, $N_{\rm ph}$, for 
which all combined sample sizes in the descent must not exceed, \textit{i.e}. $\sum_{n} \vert S_{n} \vert \leq N_{\rm ph}$. Once the total photon budget is expended we 
terminate the descent. This is to emulate an imposed restriction on computational resources required to reach a solution. \Changes{Whilst each iteration (involving forward and 
adjoint runs of the Monte Carlo) has a non-zero computational overhead, optimization of these Monte Carlo programs for repeated iteration (such as employed in~\cite{hochuli2016quantitative}) allow this overhead to become negligibly small. This means that the required computational resources of the inversion (and therefore required computation time) are proportional to the total number of simulated photons used througout the descent, \textit{i.e.} the photon budget $N_{\rm ph}$.} 
The inversions are carried out using Algorithm~\ref{alg:adaptive}, along with 
Algorithms~\ref{alg:QPAT} \& \ref{alg:UMOT} to compute the gradients for QPAT and UMOT, respectively. In Algorithm~\ref{alg:adaptive}, we will compute the metrics $V^{2}_{\rm tot}$ and $V^{2}_{\parallel}$ and conduct the norm test and inner 
product test once every 10 iterations to evaluate the quality of our computed 
gradients (using $N_{\rm rep} = 100$ independent repeated samples of the gradient), and to update the step size and sample size. Note that as this is an investigation of how such methods might 
perform in best case scenarios, we do not include the photons used to compute 
these metrics as counting against the total allowed photon budget.

\begin{table}[ht]
\centering
\begin{tabular}{c|c|c}
Strategy & Step Size, $\alpha_{n} $ & Sample Size, $\vert S_{n+1} \vert = \kappa(n)\vert S_{n} \vert$ \\ \hline
 & & \\
1        &  $\frac{1}{(1+\gamma_{\rm tot}^{2})L}$     &    $\vert S_{n+1}\vert = \frac{V_{\rm tot}^{2}}{\gamma_{\rm tot}^{2}}\vert S_{n}\vert$ \\              
 & & \\ \hline
 & & \\
2        &  $\frac{1}{(1+V_{\rm tot}^{2})L}$        &         $\vert S_{n+1}\vert = \frac{V_{\parallel}^{2}}{\gamma_{\parallel}^{2}}\vert S_{n}\vert$   \\ 
& & \\ \hline
 & & \\
3        &  $\frac{1}{(1+V_{\rm tot})L}$        &         $\vert S_{n+1}\vert = \frac{V_{\parallel}}{\gamma_{\parallel}}\vert S_{n}\vert$   \\ 

\end{tabular}
\caption{\small{Table showing the different inversion strategies used. Strategy 1 has a constant step size, with adaptive sample size. Strategies 2 \& 3 both have adaptive step sizes, and adaptive sample sizes. Note that in accordance with Algorithm~\ref{alg:adaptive}, the sample size is only increased upon a failure of the relevant test. If the test passes, then $\vert S_{n+1} \vert =  \vert S_{n} \vert$.}}
\label{tab:strats}
\end{table}

There are three different strategies we have employed to control the step size 
and sample size as the inversion progresses, see Table~\ref{tab:strats} for a 
summary. Strategy 1 uses a fixed step size as described in \eref{eqn:step} for a chosen value of $\gamma_{\rm tot}$. The sample size is adaptive and attempts to 
enforce successful outcomes of the norm test ($V^{2}_{\rm tot} \leq \gamma_{\rm tot}^{2}$), by increasing the sample size when the norm test is violated. In the 
event of a violation of this inequality, the fractional increase in the sample 
size is equivalent to the factor by which the norm test fails, $V^{2}_{\rm tot}/\gamma_{\rm tot}^{2}$. Strategy 2 uses an adaptive step size which still 
satisfies \eref{eqn:step}, however it selects the largest step size possible for 
this criteria each time the metrics are evaluated. In this strategy, the sample size is also adaptive and attempts to enforce successful outcomes of the inner product test ($V^{2}_{\parallel} \leq \gamma_{\parallel}^{2}$) by increasing the sample size 
when the inner product test is violated. In the event of a violation of this 
inequality, the fractional increase in the sample size is equivalent to the 
factor by which the inner product test fails, $V^{2}_{\parallel}/\gamma_{\parallel}^{2}$. In Strategy 3, we attempt to 
accelerate the descent by using a larger adaptive step size with $V_{\rm tot}$ in the 
denominator in place of $V^{2}_{\rm tot}$. Upon failure of the inner product 
test, the sample size is increased by fraction $V_{\parallel}/\gamma_{\parallel}$, and differs from Strategy 2 in order to reduce the speed at which 
the photon budget is depleted. This is an attempt to reduce premature increase 
of the sample size caused by volatility in the computation of the norm and inner product metrics.

Finally, we introduce an error function for the estimated absorption distribution, $\mu_{\rm a}$ 
\begin{equation}
    \Obj_{\mu_{\rm a}} = \frac{1}{2} \norm{ \mu_{\rm a}^{\rm true} - \mu_{\rm a} }^{2}.
\end{equation}
This metric would not be available under normal circumstances (as we wouldn't know the ground truth $\mu_{\rm a}^{\rm true}$), however it is useful to monitor in terms of the underlying performance of each strategy. Furthermore, as we will see, the sampled data cost function $\Obj_{S_{n}}$ is itself heavily dependent on the number of photons (sample size) used in the forward Monte Carlo, and is thus not an ideal indicator of proximity to the true solution.

\subsection{\ac{QPAT}}
\begin{figure}[ht]
\centering
\includegraphics[width=13.5cm]{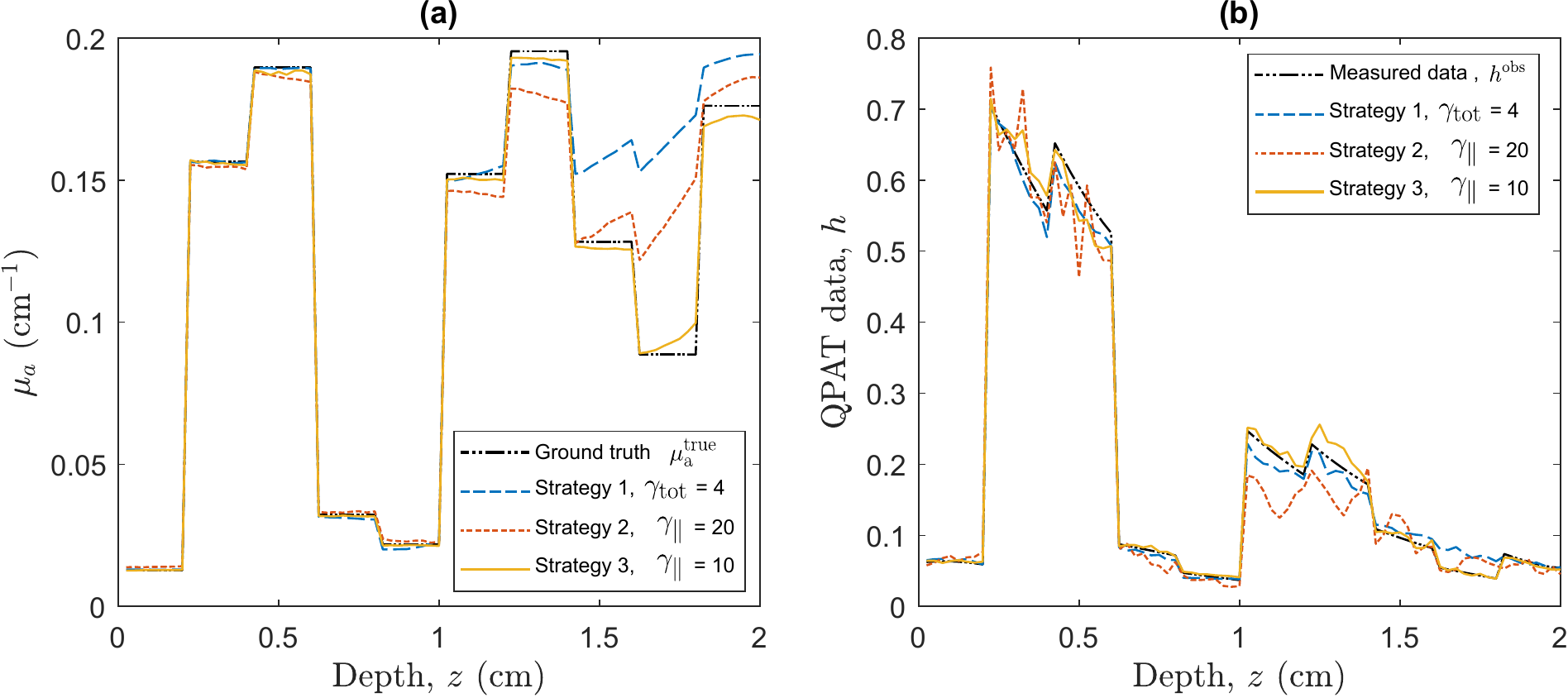}
\caption{\small{QPAT inversion: (a) - Ground truth absorption distribution, $\mu_{\rm a}^{\rm true}$, and estimated absorption distribution $\mu_{\rm a}$ at the point where the photon budget is expended, using each of the three strategies with the stated values of $\gamma_{\rm tot}$ or $\gamma_{\parallel}$. (b) - Associated measured data from ground truth medium, and simulated forward data at the end of the inversion using each strategy.}}
\label{fig:QPAT_combo_mua}
\end{figure}

\begin{figure}[ht]
\centering
\includegraphics[width=13.5cm]{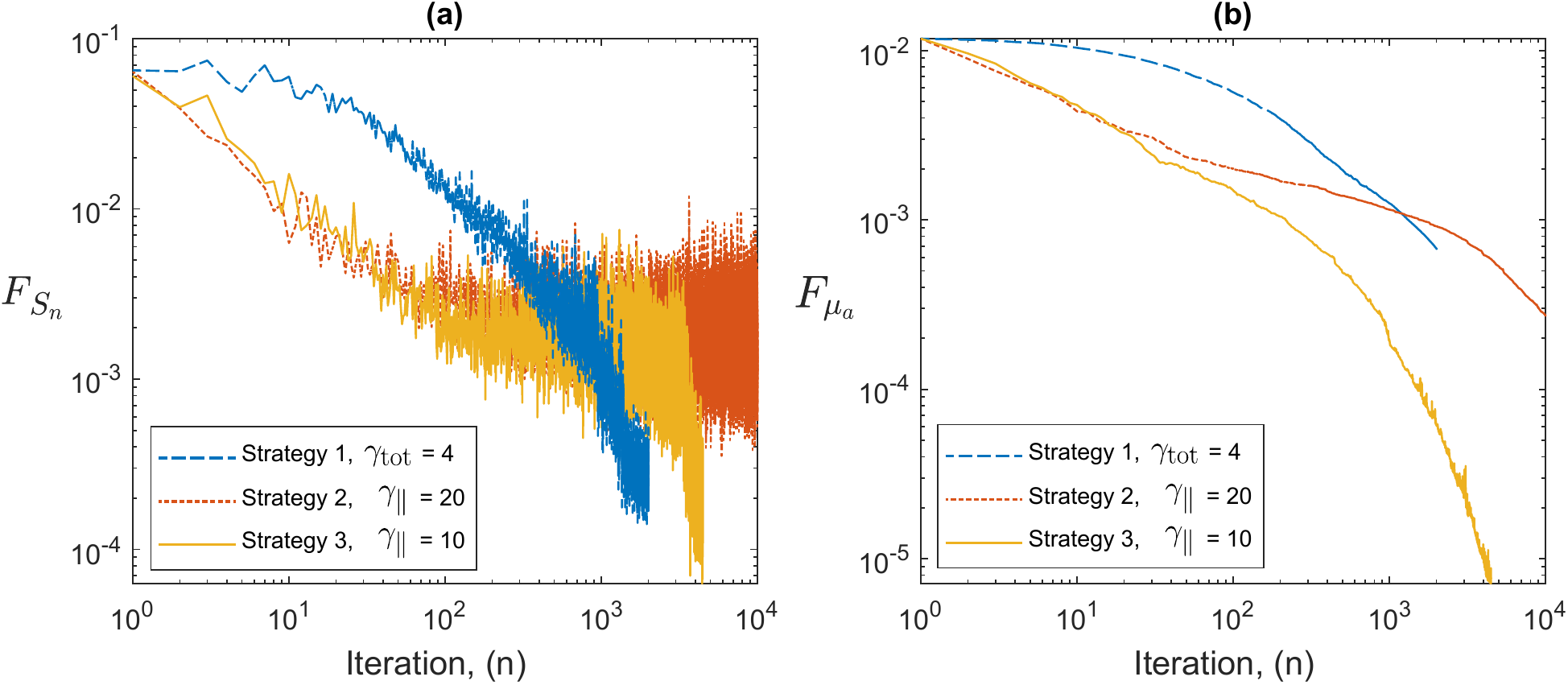}
\caption{\small{QPAT inversion: (a) - Sampled cost function, $\Obj_{S_{n}}$, as a function of iteration, $n$. (b) - Error in absorption estimate, $\Obj_{\mu_{\rm a}}$, as a function of iteration, $n$.}}
\label{fig:QPAT_combo_cost}
\end{figure}

\begin{figure}[ht]
\centering
\includegraphics[width=13.5cm]{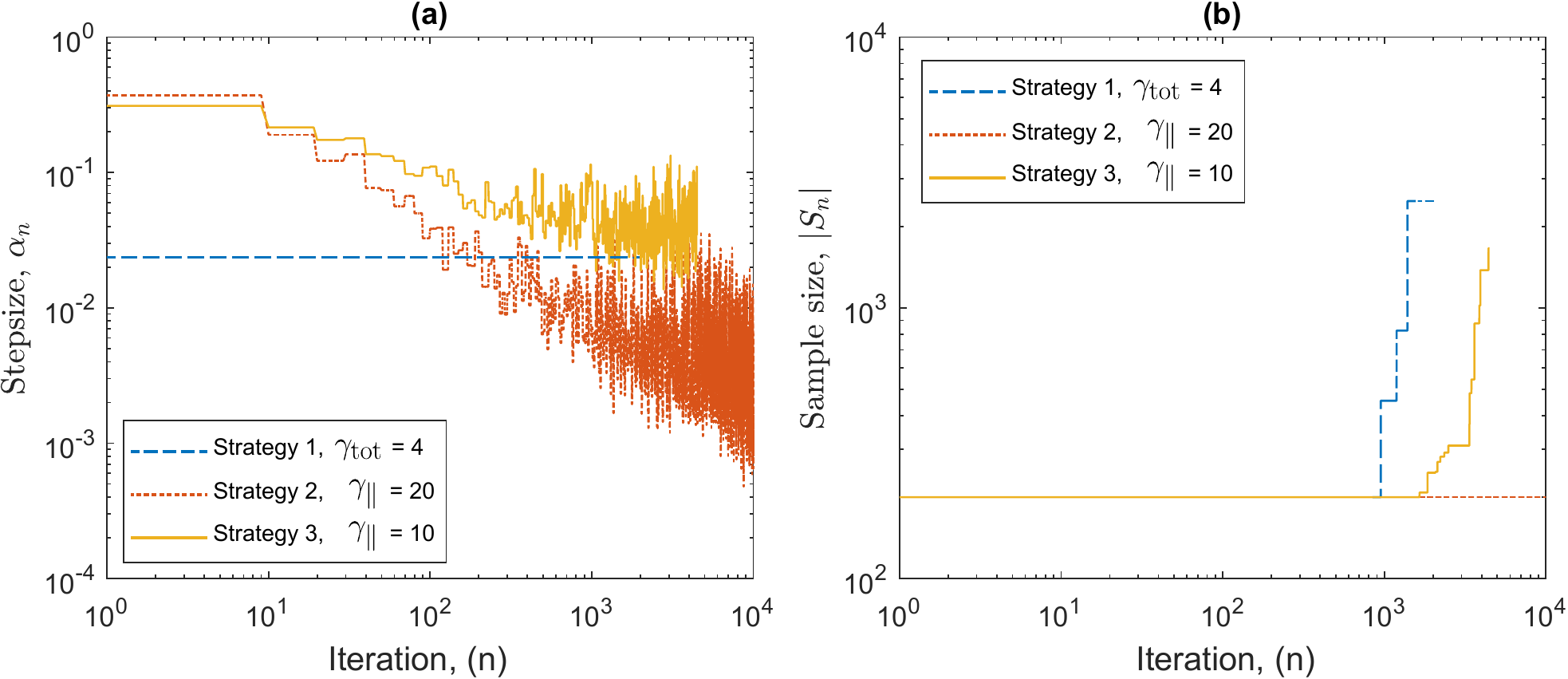}
\caption{\small{QPAT inversion: (a) - Step sizes, $\alpha_{n}$, as a function of iteration, $n$. (b) - Adaptive sample size, $\vert S_{n} \vert$, as a function of iteration.}}
\label{fig:QPAT_combo_budgets}
\end{figure}

\begin{figure}[ht]
\centering
\includegraphics[width=13.2cm]{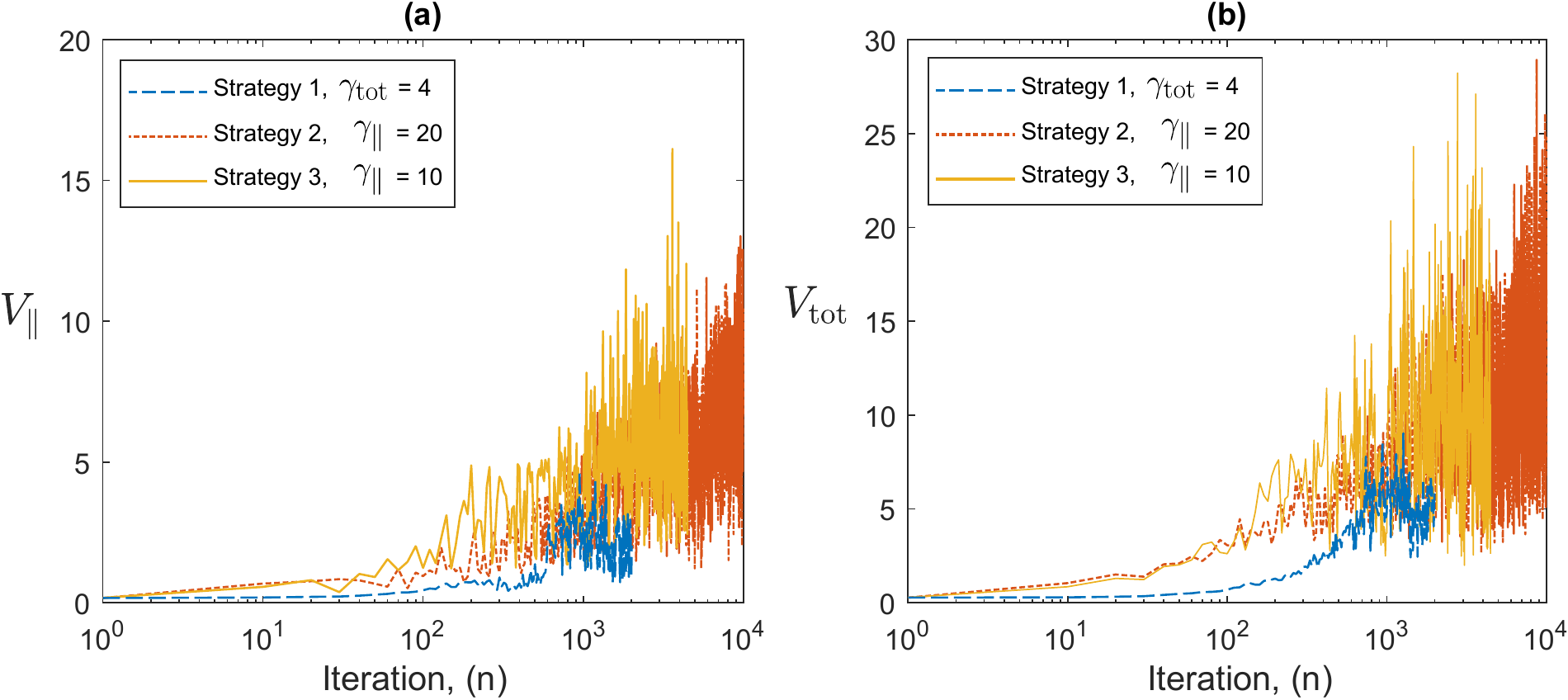}
\caption{\small{QPAT inversion: (a) - $V_{\parallel}$ as a function of iteration. (b) - $V_{\rm tot}$ as a function of iteration.}}
\label{fig:QPAT_combo_metrics}
\end{figure}

\begin{figure}[ht]
\centering
\includegraphics[width=13.8cm]{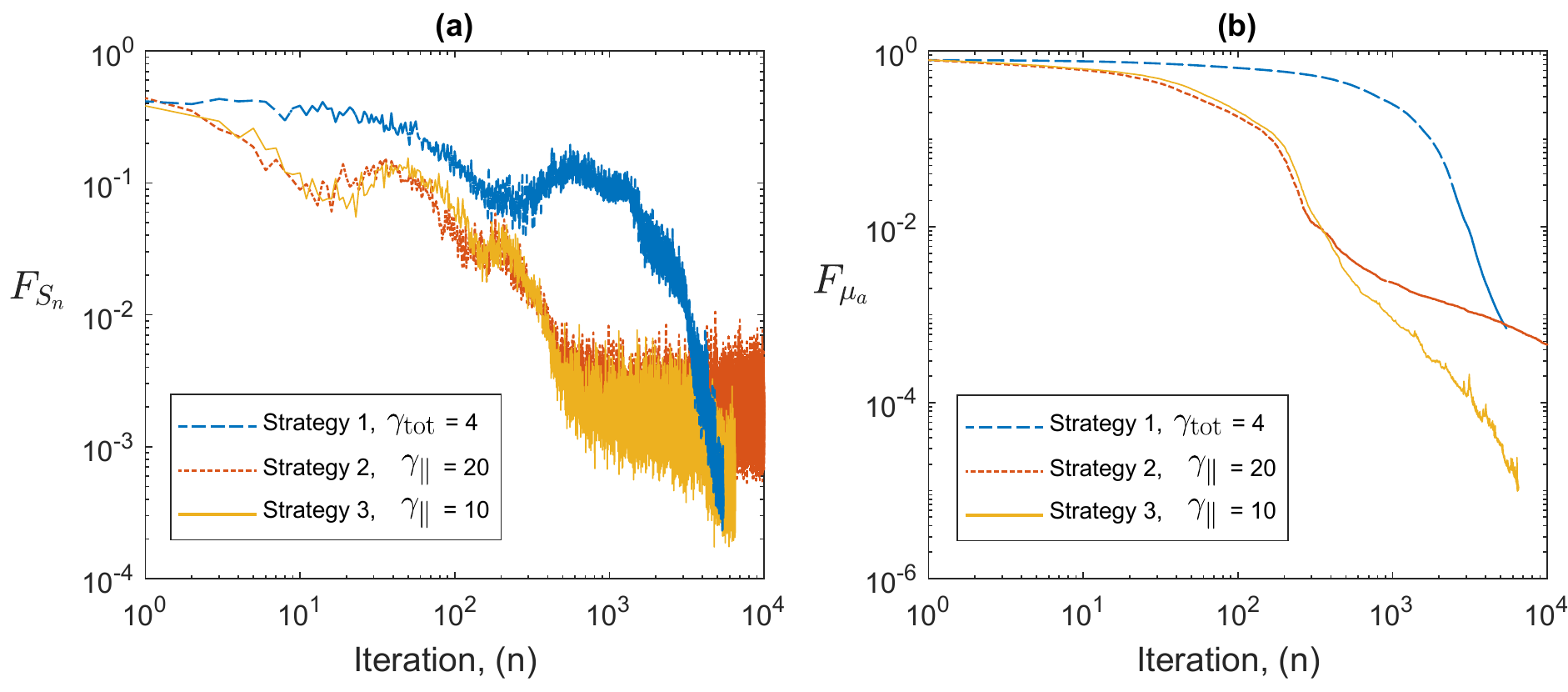}
\caption{\small{\Changes{QPAT inversion: With initial estimate of $\mua = 1.0$cm$^{-1}$: (a) - Sampled cost function, $\Obj_{S_{n}}$, as a function of iteration, $n$. (b) - Error in Absorption estimate, $\Obj_{\mu_{\rm a}}$, as a function of iteration, $n$.}}}
\label{fig:QPAT_high_mua}
\end{figure}
We begin with our example QPAT problem. The starting sample size in all cases shown is $\vert S_{1} \vert = 200$ photons per iteration (100 for each forward run, and 100 for each adjoint run in accordance with Algorithm~\ref{alg:QPAT}), and the total photon budget for the inversion was set to $N_{\rm ph} = 2 \times 10^{6}$ photons. The Lipschitz constant was set at \Changes{$L = 2.5$}, as this displayed stable descent in our test problems using large photon 
budgets (low variance case). The initial estimate of the absorption distribution in the medium is $\mu_{\rm a} = 0.2$cm$^{-1}$ in all layers. The scattering coefficient of all layers was set to \Changes{$\mus = 40$cm$^{-1}$}, and the scattering asymmetry parameter was set to \Changes{$g = 0.9$}. The ground truth absorption $\mu_{\rm a}^{\rm true}$ is shown in Fig.~\ref{fig:QPAT_combo_mua}(a), along with the final 
retrieved absorption distributions obtained via Strategies 1, 2, \& 3 using the stated values of $\gamma_{\rm tot}$ and $\gamma_{\parallel}$. Figure~\ref{fig:QPAT_combo_mua}(b) shows the corresponding ``measured'' data, and the final forward modelled data for each of the Strategies. 
Figure~\ref{fig:QPAT_combo_cost} shows the outcome of each strategy in terms of the sampled data cost function $\Obj_{S_{n}}$, and the absorption error function $\Obj_{\mu_{\rm a}}$. It can be seen that the ranking of these methods in terms of the lowest achieved value of the sampled cost function $\Obj_{S_{n}}$ does not correlate directly to the best 
outcomes in terms of the error in the estimated absorption $\Obj_{\mu_{\rm a}}$. This is due to the above mentioned dependence of the sampled data cost function on the sample size used in the forward model, where for example the case of Strategy 2 only appears to perform poorly in terms of $\Obj_{S_{n}}$ due to its small sample size used throughout the inversion. This is more clearly illustrated in Fig.~\ref{fig:QPAT_combo_mua}(b), where the final forward modelled data from Strategy 2 is noisier than the other strategies due to the low sample size \Changes{at the end of the inversion}, where this noise would clearly impact the sampled cost function. Note that the relevant step sizes, and sample sizes for each of these three examples are shown in Figure~\ref{fig:QPAT_combo_budgets}.
Before finding the best parameter for Strategy 1, we trialled a range of values of $\gamma_{\rm tot}$ over the range $[0.1,\, 20]$. With lower values, the adaptive sample size was required to increase rapidly in order to maintain high quality (low variance) sample
gradients. This resulted in the photon budget being depleted early, terminating the descent after around 100 iterations, which did not perform well. Too large a
value of $\gamma_{\parallel}$ and the norm test never failed, meaning the sample size was never required to increase and the inversion progressed for the maximum 
10,000 iterations permitted by the photon budget. However, as Strategy 1 has a fixed value of $\gamma_{\rm tot}^{2}$ in the denominator of the step size, large values also result in step sizes which were too small to perform well. A value of $\gamma_{\rm tot} = 4$ was found to strike a balance between these two extremes, and was the best performer using Strategy 1. Strategy 2 has an adaptive step size which selects the largest possible step size that still satisfies \eref{eqn:step}, instead of selecting a constant step size which accounts for the worst case scenario as in Strategy 1. For this reason, we found that the largest value of $\gamma_{\parallel} = 20$ was the best performer for this strategy, where the photon budget remained at 200 photons for each of the 10,000 iterations. For Strategy 3, the best performer was a value of \Changes{$\gamma_{\parallel} = 10$}, where 
larger values appeared to allow too much variance in the gradient, leading to unstable descents. 
In all strategies the recovered absorption distribution matched the ground truth absorption more closely in the regions of the sample closest to the light source at $z = 0$. This is due to the decay of the fluence as a function of depth, as we can see the QPAT signal is highest at shallow depths in Fig~\ref{fig:QPAT_combo_mua}(b). The \Changes{deeper} regions of the sample were the last to approach the ground truth in each of the three strategies. 

\Changes{Next}, looking at Figure~\ref{fig:QPAT_combo_metrics}, we see the values of our two metrics $V_{\parallel}$, and $V_{\rm tot}$. In all cases both measures of the variance begin at low values, indicating that even with low numbers of photons being simulated, the computed gradients are of reasonable quality, likely due to the poor initial first guess being far from the true solution. Each of the measures of variance increase as the inversion progresses until they begin to violate the norm test or inner product test depending on the strategy. It is seen that the Strategy 1 example attempts to keep $V_{\rm tot} \leq 4$, however due to some level of variation in the metrics themselves, this condition can be seen to be violated regularly, requiring regular updates to the sample size. For Strategy 2, the imposed limit of $V_{\parallel} \leq 20$ is never violated, and thus the sample size is never required to increased. We also see that Strategy 3 manages to keep \Changes{$V_{\parallel} \leq 10$} for the majority of the descent.

\Changes{In addition to these experiments presented in Figures \ref{fig:QPAT_combo_mua}-\ref{fig:QPAT_combo_metrics}, we also trialled a number of other conditions including media with isotropic scatterers (\textit{i.e} with $g= 0$), various scattering coefficients, and 
various initial estimates of the absorption. In all cases explored the 
methods showed similar behaviour as above, but with some differences 
in the ideal values of $\gamma_{\rm tot}$ and $\gamma_{\parallel}$ for 
each strategy. The outcomes of a range of these experiments are 
summarized in Table~\ref{tab:QPAT_various} for various problem 
parameters. Strategy 3 was used in all cases in the table, with the 
same Lipschitz constant ($L = 2.5$), starting sample size ($\vert S_{1} \vert$ = 200 photons), photon budget ($N_{\rm ph} = 2 \times10^{6}$ photons), and ground truth absorption distribution $\mua^{\rm true}$ as used in the above examples. The final attained 
values of the sampled data cost function $\Obj_{S_{n}}$ and absorption 
error $\Obj_{\mua}$ are similar in all cases with the exception of the 
high asymmetry and low scattering case ($g= 0.9$, and $\mus = 4$cm$^{-1}$). In this case the reduced scattering coefficient is only $\mus' = \mus(1-g) = 0.4$cm$^{-1}$, meaning much lower overall 
attenuation of the light through the sample. This results in a more 
uniform data function, $\qdata$, where the simulated photons probe the 
domain more uniformly, and allows the problem to converge significantly faster than in the higher attenuating cases demonstrated in Figs.~\ref{fig:QPAT_combo_mua}-\ref{fig:QPAT_combo_metrics}. 
It is also worth noting that in the regime with low scattering, and high scattering asymmetry, it is generally problematic for the performance of approximate transport models such as the diffusion approximation, and the results here highlight the flexibility of RTE based approaches, as well as the efficiency of the proposed adaptive sampling techniques.}

\Changes{Finally, interesting behaviour was observed when using certain initial guesses of the absorption. An example of this is shown in Fig.~\ref{fig:QPAT_high_mua}, where we show the resulting cost functions for a starting estimate of  $\mua = 1$cm$^{-1}$ (significantly overestimating the absorption at all depths), and medium properties of $g = 0.9$ and $\mu_{s} = 40$cm$^{-1}$. In this case we see that the descent appears to encounter local minima in the data cost function $\Obj_{S_{n}}$ at various points during the descent, depending on the particular strategy used. However, the algorithm manages to escape these local minima and converge to a better solution. This is seen to be the case for all three strategies shown in Fig.~\ref{fig:QPAT_high_mua}.}

\begin{table}[]
\begin{tabular}{ccccc}
                                                            &                                                                                                  &                                                                                                                                                                                   & Starting $\mu_{a}$ (cm$^{-1}$)                                                                                                                                                    &                                                                                                                                                              \\
\multicolumn{1}{l}{}                                        & \multicolumn{1}{l}{}                                                                             & \multicolumn{1}{l}{}                                                                                                                                                              & \multicolumn{1}{l}{}                                                                                                                                                              & \multicolumn{1}{l}{}                                                                                                                                         \\
                                                            & \multicolumn{1}{c|}{}                                                                            & \multicolumn{1}{c|}{0.01}                                                                                                                                                         & \multicolumn{1}{c|}{0.2}                                                                                                                                                          & 1.0                                                                                                                                                          \\ \cline{2-5} 
                                                            & \multicolumn{1}{c|}{\begin{tabular}[c]{@{}c@{}}$g = 0.9$\\ $\mu_{s} = 40$cm$^{-1}$\end{tabular}} & \multicolumn{1}{c|}{\begin{tabular}[c]{@{}c@{}}$\gamma_{\parallel} = 20$\\ 10000 iterations\\ $F_{S_{n}} = 2.35\times 10^{-3}$\\ $F_{\mu_{a}} = 3.26\times 10^{-5}$\end{tabular}} & \multicolumn{1}{c|}{\begin{tabular}[c]{@{}c@{}}$\gamma_{\parallel} = 10$\\ 4476 iterations\\ $F_{S_{n}} = 4.20\times 10^{-4}$\\ $F_{\mu_{a}} = 7.65\times 10^{-6}$\end{tabular}}  & \begin{tabular}[c]{@{}c@{}}$\gamma_{\parallel} = 10$\\ 6533 iterations\\ $F_{S_{n}} = 3.38\times 10^{-4}$\\ $F_{\mu_{a}} = 1.01\times 10^{-5}$\end{tabular}  \\ \cline{2-5} 
\begin{tabular}[c]{@{}c@{}}Medium\\ Properties\end{tabular} & \multicolumn{1}{c|}{\begin{tabular}[c]{@{}c@{}}$g = 0.9$\\ $\mu_{s} = 4$cm$^{-1}$\end{tabular}}  & \multicolumn{1}{c|}{\begin{tabular}[c]{@{}c@{}}$\gamma_{\parallel} = 5$\\ 3819 iterations\\ $F_{S_{n}} = 5.30\times 10^{-4}$\\ $F_{\mu_{a}} = 2.3\times 10 ^{-7}$\end{tabular}}   & \multicolumn{1}{c|}{\begin{tabular}[c]{@{}c@{}}$\gamma_{\parallel} = 5$\\ 2579 iterations\\ $F_{S_{n}} = 9.31 \times 10^{-5}$\\ $F_{\mu_{a}} = 3.56 \times 10^{-7}$\end{tabular}} & \begin{tabular}[c]{@{}c@{}}$\gamma_{\parallel} = 5$\\ 2834 iterations\\ $F_{S_{n}} = 1.06\times 10^{-4}$\\ $F_{\mu_{a}} = 2.19\times 10^{-7}$\end{tabular}   \\ \cline{2-5} 
                                                            & \multicolumn{1}{c|}{\begin{tabular}[c]{@{}c@{}}$g = 0$\\ $\mu_{s} = 4$cm$^{-1}$\end{tabular}}    & \multicolumn{1}{c|}{\begin{tabular}[c]{@{}c@{}}$\gamma_{\parallel} = 20$\\ 10000 iterations\\ $F_{S_{n}} = 4.01\times 10^{-3}$\\ $F_{\mu_{a}} = 8.36\times 10^{-5}$\end{tabular}} & \multicolumn{1}{c|}{\begin{tabular}[c]{@{}c@{}}$\gamma_{\parallel} = 5$\\ 2056 iterations\\ $F_{S_{n}} = 1.39\times 10^{-4}$\\ $F_{\mu_{a}} = 3.44\times 10^{-5}$\end{tabular}}   & \begin{tabular}[c]{@{}c@{}}$\gamma_{\parallel} = 10$\\ 10000 iterations\\ $F_{S_{n}} = 2.92\times 10^{-3}$\\ $F_{\mu_{a}} = 8.72\times 10^{-5}$\end{tabular}
\end{tabular}
\caption{
\Changes{\small{Final outcomes of QPAT inversions with various medium optical properties and starting values of $\mua$. Values of $\Obj_{S_{n}}$ and $\Obj_{\mua}$ are the final values at the end of each inversion after the stated number of iterations. In each case Strategy 3 was employed, with a starting sample size of $\vert S_{1} \vert = 200$ photons per iteration, and a total photon budget of $N_{\rm ph} = 2\times 10^{6}$ photons. Slab thickness is $2$cm in all cases, with the same ground truth $\mua^{\rm true}$ distribution as shown in Fig.~\ref{fig:QPAT_combo_mua}(a).}}
} 
\label{tab:QPAT_various}
\end{table}

\subsection{UMOT}
\begin{figure}[ht]
\centering
\includegraphics[width=13.5cm]{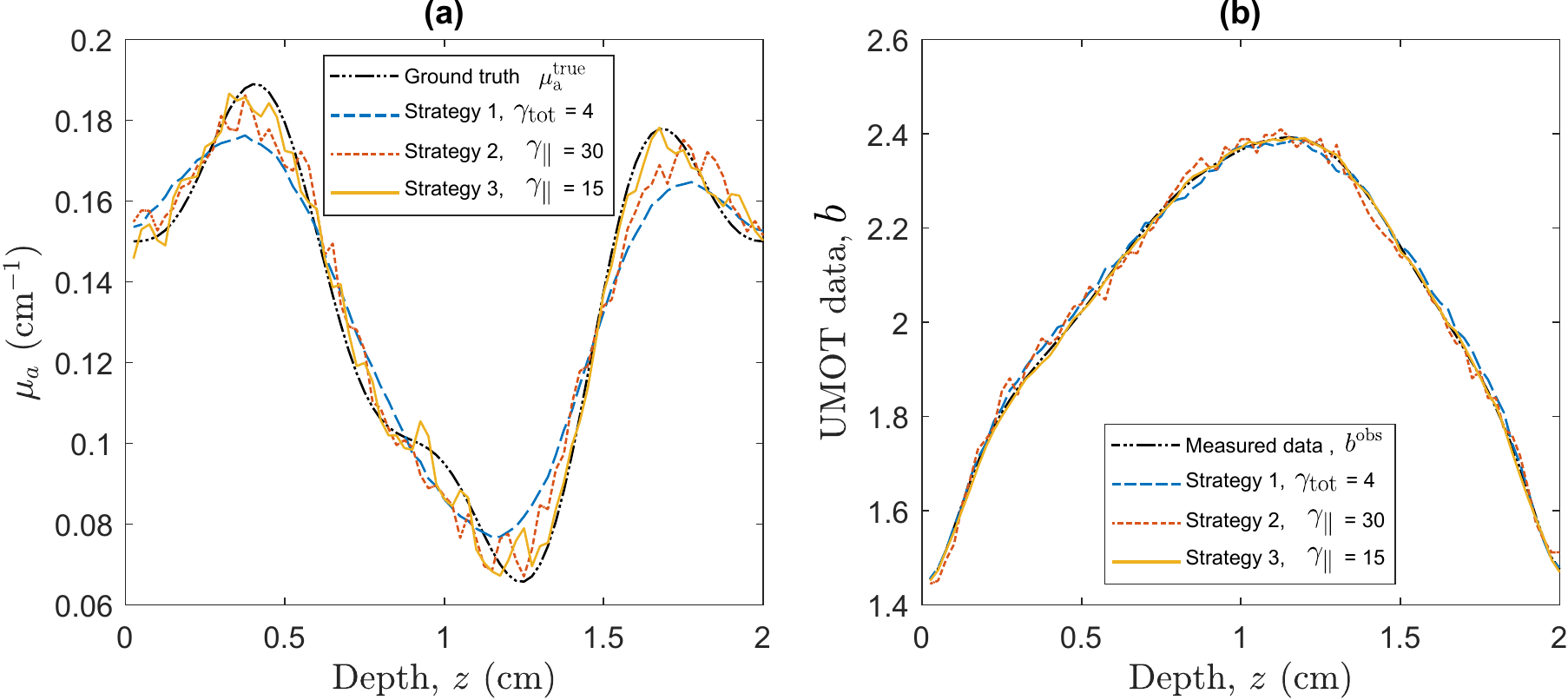}
\caption{\small{UMOT inversion: (a) - Ground truth absorption distribution, $\mu_{\rm a}^{\rm true}$, and recovered absorption distribution $\mu_{\rm a}$ using each of the three strategies with the stated values of $\gamma_{\rm tot}$ or $\gamma_{\parallel}$. (b) - Associated measured data from ground truth medium, and simulated forward data at the end of the inversion using each strategy.}}
\label{fig:UMOT_combo_mua}
\end{figure}
\begin{figure}[ht]
\centering
\includegraphics[width=14cm]{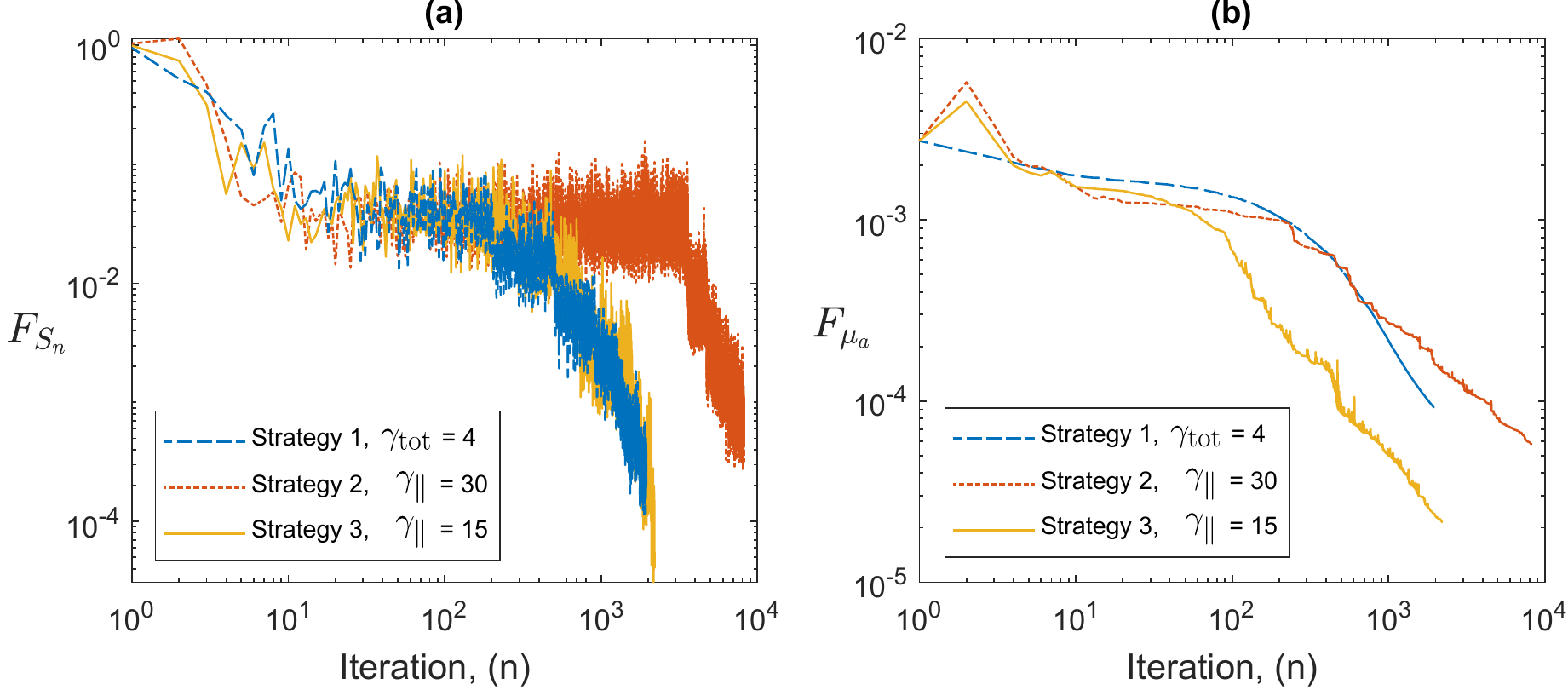}
\caption{\small{UMOT inversion: (a) - Sampled cost function, $\Obj_{S_{n}}$, as a function of iteration, $n$. (b) - Error in absorption estimate, $\Obj_{\mu_{\rm a}}$, as a function of iteration, $n$.}}
\label{fig:UMOT_combo_cost}
\end{figure}
\begin{figure}[ht]
\centering
\includegraphics[width=14cm]{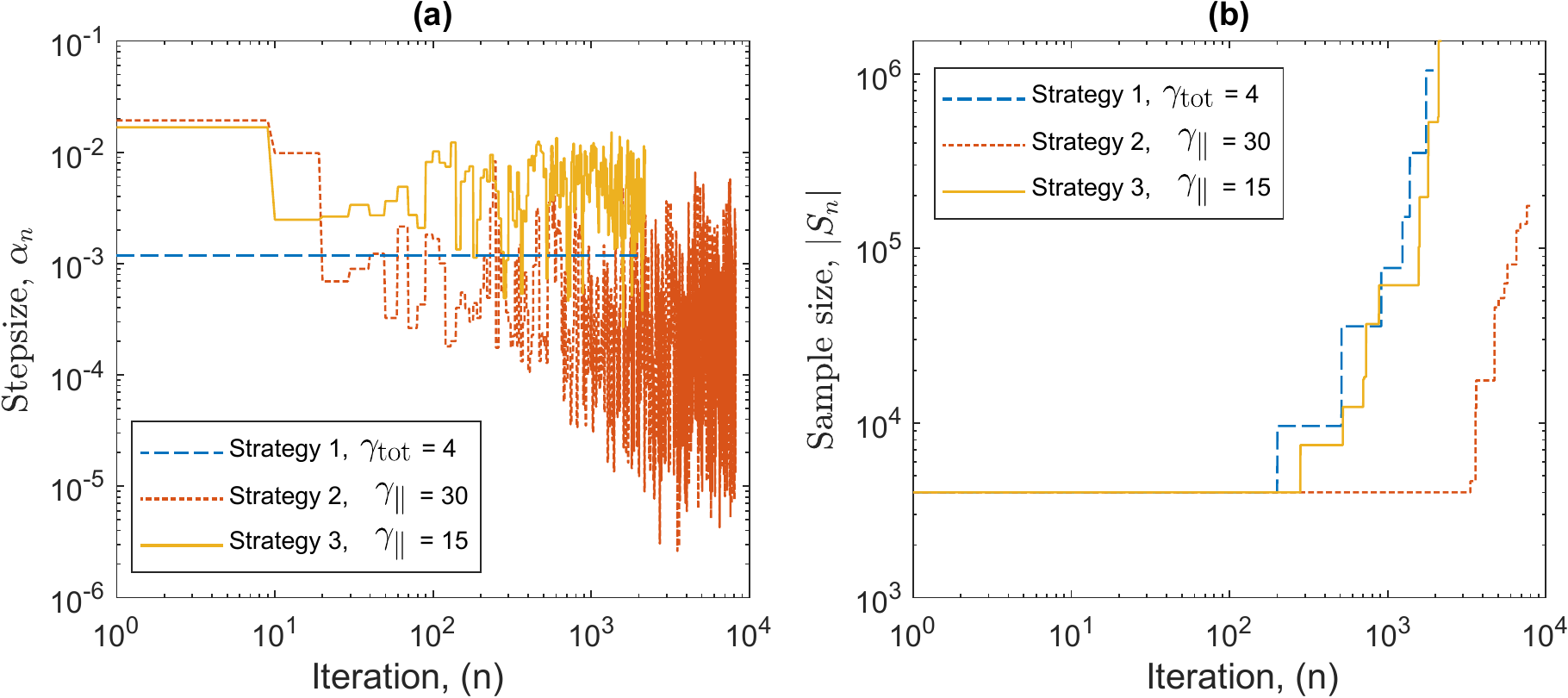}
\caption{\small{UMOT inversion: (a) - Step sizes, $\alpha_{n}$, as a function of iteration, $n$. (b) - Adaptive sample size, $\vert S_{n} \vert$, as a function of iteration.}}
\label{fig:UMOT_combo_budgets}
\end{figure}
\begin{figure}[ht]
\centering
\includegraphics[width=14cm]{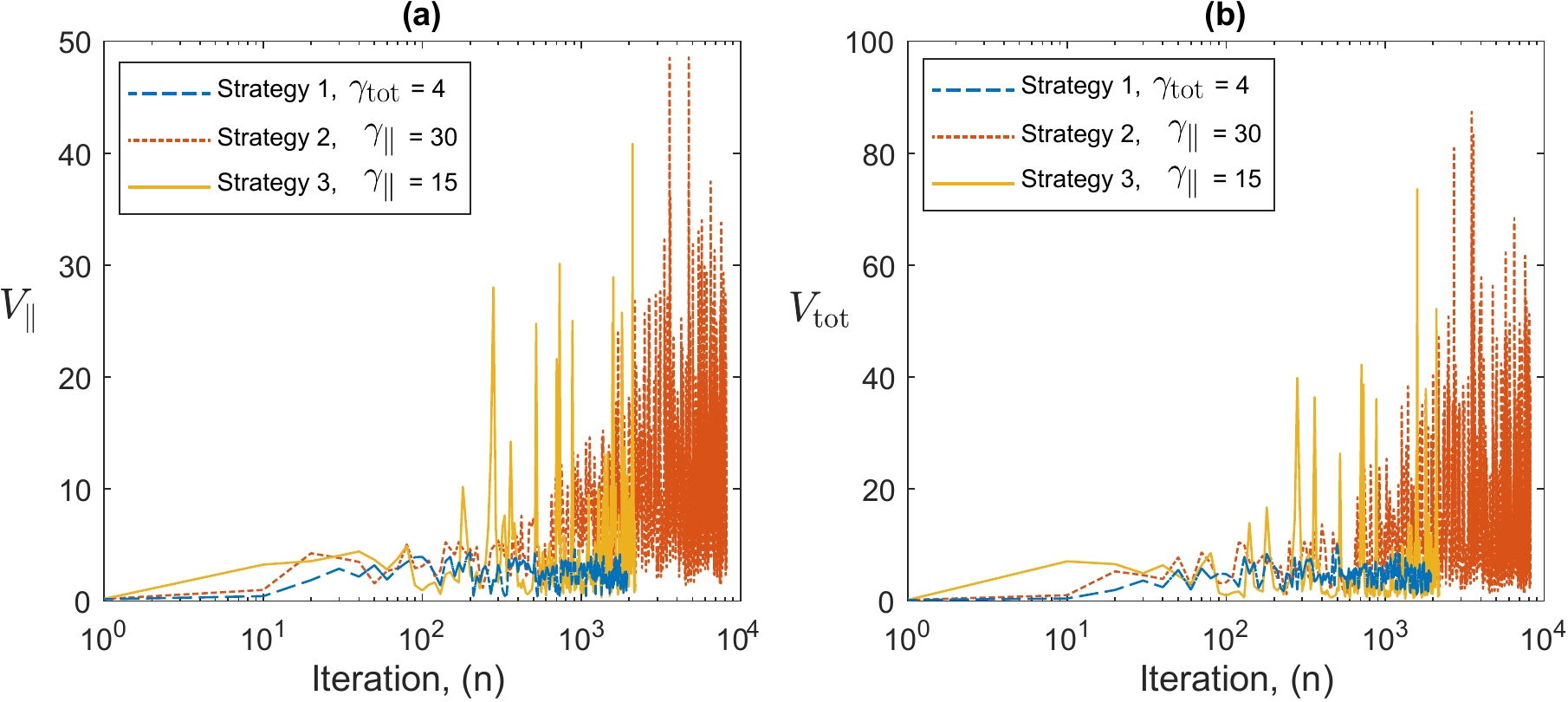}
\caption{\small{UMOT inversion: (a) - $V_{\parallel}$ as a function of iteration. (b) - $V_{\rm tot}$ as a function of iteration.}}
\label{fig:UMOT_combo_metrics}
\end{figure}

Next we demonstrate similar experiments performed using the UMOT modality 
described in Section~\ref{sec:UMOT_case} for the transmission geometry. In this 
setup we used the same medium slab size as the QPAT example, and the same optical properties apart from the absorption distribution.
The starting sample size in all cases shown is $\vert S_{1} \vert = 4000$ photons per iteration, 1000 for each forward run (per each of the two sources), and 1000 
for each of the two adjoint sources as outlined in Algorithm~\ref{alg:UMOT}. The total photon budget for the inversion was set to $N_{\rm ph} = 4 \times 10^{8}$ photons. The Lipschitz constant was set at \Changes{$L = 50$}, as this 
displayed stable descent in our test problems using large photon budgets (low 
variance case). The initial estimate of the absorption distribution in the medium is $\mu_{\rm a} = 0.1$cm$^{-1}$ in all layers. The ground truth absorption $\mu_{\rm a}^{\rm true}$ is shown in Fig.~\ref{fig:UMOT_combo_mua}(a), along with the final retrieved absorption distributions obtained via Strategies 1, 2, \& 3 using the 
stated values of $\gamma_{\rm tot}$ and $\gamma_{\parallel}$. Figure~\ref{fig:UMOT_combo_mua}(b) shows the true ``measured'' UMOT data, $b^{\rm obs}$, along with the forward modelled data from the final estimated medium for each strategy.
Figure~\ref{fig:UMOT_combo_cost} shows the outcome of each strategy in terms of 
the sampled data cost function $\Obj_{S_{n}}$, and the absorption error function $\Obj_{\mu_{\rm a}}$. The relevant step sizes, and sample sizes for each of these 
three examples are shown in Figure~\ref{fig:UMOT_combo_budgets}, and the values 
of the metrics measuring the variance in the sampled gradients are presented in 
Figure~\ref{fig:UMOT_combo_metrics}. Similar to the the QPAT modality, we found 
that Strategy 3 performed the best in terms of the final achieved value of the 
error in the absorption estimate $\Obj_{\mu_{\rm a}}$.

\Changes{In addition to the results shown in Figures~\ref{fig:UMOT_combo_mua}-\ref{fig:UMOT_combo_metrics}, in Table~\ref{tab:UMOT_various} we also present a summary of results for a range 
of different medium optical parameters, and starting estimates of the 
absorption. In all cases, Strategy 3 was used, and the starting photon 
budget was the same as in the previous \ac{UMOT} examples ($\vert S_{1} \vert = 4000$ photons), with a total photon budget of $N_{\rm ph} = 4\times 10^{8}$ photons. For each of the inversions presented in this 
table, we conducted the inner product test once every 50 iterations, 
using $N_{\rm rep} = 50$ repeated evaluations of the gradient. The 
resulting inversions display similar error in these cases to the above 
examples where we used $N_{\rm rep} = 100$ repeated evaluations of the 
sampled gradient once every 10 iterations to run the inner product test. 
This demonstrates that the described methods can still be successful 
when dedicating fewer computational resources to the inner product or 
norm test metrics which control the adaptive sample size, and step size.
}

\begin{table}[]
\begin{tabular}{ccccc}
                                                            &                                                                                                  &                                                                                                                                                                                    & Starting $\mu_{a}$ (cm$^{-1}$)                                                                                                                                                      &                                                                                                                                                              \\
\multicolumn{1}{l}{}                                        & \multicolumn{1}{l}{}                                                                             & \multicolumn{1}{l}{}                                                                                                                                                               & \multicolumn{1}{l}{}                                                                                                                                                                & \multicolumn{1}{l}{}                                                                                                                                         \\
                                                            & \multicolumn{1}{c|}{}                                                                            & \multicolumn{1}{c|}{0.01}                                                                                                                                                          & \multicolumn{1}{c|}{0.1}                                                                                                                                                            & 1.0                                                                                                                                                          \\ \cline{2-5} 
                                                            & \multicolumn{1}{c|}{\begin{tabular}[c]{@{}c@{}}$g = 0.9$\\ $\mu_{s} = 40$cm$^{-1}$\end{tabular}} & \multicolumn{1}{c|}{\begin{tabular}[c]{@{}c@{}}$\gamma_{\parallel} = 15$\\ 11412 iterations\\ $F_{S_{n}} = 1.88\times 10^{-3}$\\ $F_{\mu_{a}} = 4.23\times 10^{-5}$\end{tabular}}  & \multicolumn{1}{c|}{\begin{tabular}[c]{@{}c@{}}$\gamma_{\parallel} = 10$\\ 3495 iterations\\ $F_{S_{n}} = 2.50\times 10^{-4}$\\ $F_{\mu_{a}} = 2.20\times 10^{-5}$\end{tabular}}    & \begin{tabular}[c]{@{}c@{}}$\gamma_{\parallel} = 10$\\ 7823 iterations\\ $F_{S_{n}} = 5.69\times 10^{-4}$\\ $F_{\mu_{a}} = 9.37\times 10^{-5}$\end{tabular}  \\ \cline{2-5} 
\begin{tabular}[c]{@{}c@{}}Medium\\ Properties\end{tabular} & \multicolumn{1}{c|}{\begin{tabular}[c]{@{}c@{}}$g = 0.9$\\ $\mu_{s} = 4$cm$^{-1}$\end{tabular}}  & \multicolumn{1}{c|}{\begin{tabular}[c]{@{}c@{}}$\gamma_{\parallel} = 15$\\ 19017 iterations\\ $F_{S_{n}} = 4.71\times 10^{-3}$\\ $F_{\mu_{a}} = 8.19\times 10 ^{-5}$\end{tabular}} & \multicolumn{1}{c|}{\begin{tabular}[c]{@{}c@{}}$\gamma_{\parallel} = 15$\\ 15813 iterations\\ $F_{S_{n}} = 5.48 \times 10^{-3}$\\ $F_{\mu_{a}} = 2.07 \times 10^{-5}$\end{tabular}} & \begin{tabular}[c]{@{}c@{}}$\gamma_{\parallel} = 15$\\ 14249 iterations\\ $F_{S_{n}} = 4.08\times 10^{-3}$\\ $F_{\mu_{a}} = 3.36\times 10^{-5}$\end{tabular} \\ \cline{2-5} 
                                                            & \multicolumn{1}{c|}{\begin{tabular}[c]{@{}c@{}}$g = 0$\\ $\mu_{s} = 4$cm$^{-1}$\end{tabular}}    & \multicolumn{1}{c|}{\begin{tabular}[c]{@{}c@{}}$\gamma_{\parallel} = 15$\\ 11594 iterations\\ $F_{S_{n}} = 1.65\times 10^{-3}$\\ $F_{\mu_{a}} = 7.13\times 10^{-5}$\end{tabular}}  & \multicolumn{1}{c|}{\begin{tabular}[c]{@{}c@{}}$\gamma_{\parallel} = 15$\\ 7304 iterations\\ $F_{S_{n}} = 5.87\times 10^{-4}$\\ $F_{\mu_{a}} = 4.17\times 10^{-5}$\end{tabular}}    & \begin{tabular}[c]{@{}c@{}}$\gamma_{\parallel} = 15$\\ 13981 iterations\\ $F_{S_{n}} = 1.05\times 10^{-3}$\\ $F_{\mu_{a}} = 7.98\times 10^{-5}$\end{tabular}
\end{tabular}
\caption{
\Changes{\small{Final outcomes of UMOT inversions with various medium optical properties and starting values of $\mua$. Values of $\Obj_{S_{n}}$ and $\Obj_{\mua}$ are the final values at the end of each inversion after the stated number of iterations. In each case Strategy 3 was employed, with a starting sample size of $\vert S_{1} \vert = 4000$ photons per iteration, and a total photon budget of $N_{\rm ph} = 4\times 10^{8}$ photons. Slab thickness is $2$cm in all cases, with the same ground truth $\mua^{\rm true}$ distribution as shown in Fig.~\ref{fig:UMOT_combo_mua}(a).}}
} 
\label{tab:UMOT_various}
\end{table}

\section{Discussion and Conclusions}
\label{sec:discussion}

The results shown in Section~\ref{sec:results} demonstrate that the adaptive sampling strategies performed well in both our example problems of \ac{QPAT} and \ac{UMOT}. We were able to achieve low error estimates of the medium absorption using a total computational expenditure that was either comparable to, or significantly lower than the resources required to simulate a single low variance run of the forward problem. In each demonstration the adaptive sampling strategies maintained low photon numbers throughout the early stages of the inversion. \Changes{Photon} numbers were only increased when required to keep the variance in the gradients below the stipulated limits. These adaptive sampling strategies thus enabled significant computational savings compared to a na\"ive implementation which might seek to use low variance (high quality) computations of the gradient at every iteration. 
\Changes{For instance, if we were to use a constant stepsize of $1/L$, and the same number of photons per iteration as that 
which was used to generate the ``measured'' data ($10^{9}$ photons), 
then we find we still required hundreds of iterations to reach a similar quality estimate of the absorption as seen in the above problems. This means that the computational requirements of the low variance approach would be proportional to $N_{\rm ph} = 10^{11}$ photons. Comparing this to $N_{\rm ph} = 2\times10^{6}$ photons used in the \ac{QPAT} examples, or $N_{\rm ph} = 4\times 10^{8}$ photons used in the \ac{UMOT} examples, the required computational resources/time to attain our solutions with these adaptive sampling methods is multiple orders of magnitude lower compared to the na\"ive low variance approach.}

In this work we have emphasised the similarities between our approach 
and that of \acl{SGD}, as employed in the context of machine learning. 
However it should be noted that there are significant differences between the two settings. In machine learning, the measured data are assumed to 
consist of a large number of samples to be fit to a \emph{deterministic} 
model so as to minimise a suitable loss function, and each stochastic 
gradient is generated by a random subset of these data forming the 
descent direction of a sub-function. The same method has also been 
applied in alternative image reconstruction techniques where the data 
can be more naturally considered as consisting of a large number of 
random samples from some underlying distribution, for example in Positron Emission Tomography~\cite{thielemans2015}. By contrast, our image 
reconstruction approach considers the complete measured data on each 
iteration, with stochasticity arising from the approximation within the 
forward model: we are effectively sub-sampling the gradient in terms of 
the parameter space, rather than data space. This is to say that at each 
iteration we utilise a subset of some notionally complete model, rather 
than of the data. The motivation by which each approach is employed is 
consistent: stochasticty is intentionally introduced to whichever part 
of the objective function introduces the greatest computational demand. 
This suggests a third possible approach, where the computational load of 
the (sub-) gradient computation can be lowered through some stochastic 
division of both the data, and the model; this might be relevant in 
imaging modalities with discrete counting data, such as time-domain 
and/or dynamic diffuse optical tomography.

Our work suggests a number of interesting future developments:

\begin{itemize}
    \item In the examples shown here the ``observed'' data were effectively ``noise-free'' by virtue of running the forward Monte Carlo on a very large number of photons. Thus an interesting topic for further study will be to evaluate these methods on noisy forward data, wherein the data fitting term should not be iterated to convergence, but where regularisation should be introduced either by early stopping (i.e. by setting a minimum threshold for the data error) or by adding an explicit penalty term. 
    \item Related to the previous point, we further note that our objective function  employed a least squares data fitting term in this study. Depending upon the nature of the noise in the data \emph{and} that of the stochastic forward model, more suitable metrics may include the Kullback-Leibler discrepency (for Poisson likelihood), or 
    a generalised measure of the distance between samples of probability distributions (Wasserstein distance\cite{villani2009}).
    \item  Our results demonstrate a consistent tendency for the adaptive sampling method to exhibit a geometric increase in the sample size as the descent progresses. This suggests that our adaptive approach could be employed to find a particular set of sampling parameters that perform well in a given regime, including the starting photon budget $\vert S_{1} \vert$, rate of increase of the sample size $\kappa(n)$, and rate of change of the step size $\alpha_{n}$. If a suitable set of such parameters could be found, they could help determine a fully \emph{prescribed} sampling strategy. Once calibrated for a given problem of choice, this would avoid the need to explicitly compute the variance of the sampled gradients during the descent, and lead to even greater efficiency and speed in the inverse problem. 
    \item Further topics of interest \Changes{include} more advanced methods of variance reduction (e.g recursive gradients~\cite{Nguyen2017})\Changes{;} adaptive estimates of the Lipschitz constant as described in Ref.~\cite{bolla2018adaptive}\Changes{;} alternative optimisation strategies such as back-tracking line-search, or primal dual methods\cite{chambolle2018}\Changes{;} the use of preconditioning and/or second-order optimisation methods\cite{moritz2016}\Changes{; and an in depth comparison of these non-linear adaptive models to the alternative approaches such as Perturbation Monte Carlo~\cite{leino2019perturbation}}.   
\end{itemize}

In summary, we have successfully demonstrated a means by which stochastic forward models, not directly amenable to standard variational methods for optimisation, can be employed efficiently in non-linear image reconstruction. We expect this concept to lead to be many new directions of research in optical image reconstruction.

\section*{Disclosures}

No conflicts of interest, financial or otherwise, are declared by the authors.

\section*{Acknowledgements}

This work was funded by Engineering and Physical Sciences Research Council (EPSRC), EP/N032055/1 \& EP/N025946/1. S. Powell further acknowledges Royal Academy of Engineering (RAEng) fellowship RF1516/15/33.

The authors would like to thank Robert Twyman and Kris Thielemans for helpful discussions on
stochastic gradient and data subset methods.

\bibliography{refs}
\bibliographystyle{spiebib}

\newpage

\end{document}